\documentclass[titlepage, twocolumn, aps, prd, longbibliography, nofootinbib, superscriptaddress, floatfix]{revtex4-1}
\pdfoutput=1

\usepackage[english]{babel}
\usepackage{appendix}
\usepackage{graphicx}
\usepackage{amsmath}
\usepackage{amsfonts}
\usepackage{amsthm}

\usepackage{apptools}

\newtheorem{lemma}{Lemma}

\newtheorem{proposition}{Proposition}

\usepackage{dsfont}
\usepackage{mathtools}
\usepackage{xcolor}
\usepackage[caption=false]{subfig}

\usepackage[pdftex,colorlinks=true]{hyperref}
\hypersetup{linkcolor=blue,citecolor = blue,urlcolor=blue}

\definecolor{MyDarkBlue}{rgb}{0,0.08,0.45} 
\definecolor{MyLightMagenta}{cmyk}{0.1,0.8,0,0.1} 
\definecolor{MyDarkGreen}{rgb}{0,0.45,0.08} 
\definecolor{lightcarminepink}{rgb}{0.9, 0.4, 0.38}
\definecolor{darkpastelpurple}{rgb}{0.59, 0.44, 0.84}
\definecolor{purple}{rgb}{0.63, 0.36, 0.94}

\newcommand{\RNum}[1]{\uppercase\expandafter{\romannumeral #1\relax}}

\newcommand\numberthis{\addtocounter{equation}{1}\tag{\theequation}}

\newcommand{\ket}[1]{| #1 \rangle}
\newcommand{\bra}[1]{\langle #1 |}

\DeclarePairedDelimiter\abs{\lvert}{\rvert}%
\DeclarePairedDelimiter\norm{\lVert}{\rVert}%
\def\smallnorm#1{ {|\hspace{-.015in}|#1|\hspace{-.015in}|} }

\makeatletter
\let\oldabs\abs
\def\abs{\@ifstar{\oldabs}{\oldabs*}}
\let\oldnorm\norm
\def\norm{\@ifstar{\oldnorm}{\oldnorm*}}
\makeatother

\AtAppendix{\counterwithin{lemma}{section}}
\AtAppendix{\counterwithin{proposition}{section}}

\begin{document}

\title{Classically computing performance bounds on depolarized quantum circuits}

\author{Sattwik Deb Mishra$^{*}$}
\affiliation{Ginzton Laboratory, Stanford University, 348 Via Pueblo Mall, Stanford, California 94305, USA} 
\author{Miguel Fr\'ias-P\'erez$^{*}$}
\affiliation{Max-Planck-Institute of Quantum Optics, Hans-Kopfermann-Str. 1, Garching 85748, Germany}
\author{Rahul Trivedi\hyperlink{email}{$^{\dagger}$}}
\affiliation{Max-Planck-Institute of Quantum Optics, Hans-Kopfermann-Str. 1, Garching 85748, Germany}
\affiliation{Electrical and Computer Engineering, University of Washington, Seattle, Washington 98195, USA}

\begin{abstract}

Quantum computers and simulators can potentially outperform classical computers in finding ground states of classical and quantum Hamiltonians. However, if this advantage can persist in the presence of noise without error correction remains unclear. In this paper, by exploiting the principle of Lagrangian duality, we develop a numerical method to classically compute a certifiable lower bound on the minimum energy attainable by the output state of a quantum circuit in the presence of depolarizing noise. We provide theoretical and numerical evidence that this approach can provide circuit-architecture dependent bounds on the performance of noisy quantum circuits.  
\end{abstract}

%\showthe\textwidth

\maketitle
\def\thefootnote{}\footnotetext{$^*$ These authors contributed equally to this work. \hypertarget{email}{$^{\dagger}$} rtriv@uw.edu}

\section{Introduction}

Fault-tolerant quantum computers hold promise for outperforming classical computers at several computational tasks. One of the most explored computational tasks is the problem of finding the ground state of a given many-body Hamiltonian --- a problem that naturally arises in studying equilibrium properties of condensed matter systems \cite{amico2008entanglement}. Moreover, classical optimization problems can also be framed as finding ground states of commuting Hamiltonians \cite{gharibian2015quantum}. Unsurprisingly, quantum algorithms for finding Hamiltonian ground states have been extensively studied \cite{verstraeteQuantumComputationQuantumstate2009a, albashAdiabaticQuantumComputation2018, mottaDeterminingEigenstatesThermal2020} in search of a possible quantum advantage \cite{zhongQuantumComputationalAdvantage2020b, kimEvidenceUtilityQuantum2023, wuStrongQuantumComputational2021} --- algorithms based on phase estimation and adiabatic evolution have been proposed for solving this problem, and have even been shown to be efficient for specific classes of Hamiltonians \cite{geFasterGroundState2019c, geRapidAdiabaticPreparation2016a}. Furthermore, due to the constraints on available quantum hardware, there has been intense activity in exploring hardware-efficient heuristics for solving this problem such as quantum adiabatic algorithms or variational quantum algorithms \cite{cerezoVariationalQuantumAlgorithms2021, farhiQuantumApproximateOptimization2014, weckerProgressPracticalQuantum2015, bhartiNoisyIntermediatescaleQuantum2022}.

Current noisy-intermediate scale quantum devices, however, do not perform quantum error correction and consequently noise places a severe constraint on the performance of these quantum algorithms. From a theoretical standpoint, it has thus become of interest to develop no-go results by providing theoretical bounds on the minimum energy that a noisy quantum circuit can achieve for a given Hamiltonian  --- if  a classical algorithm  \cite{goemansImprovedApproximationAlgorithms1995, anjos2004semidefinite, oliveira2005complexity} could obtain an energy better than this lower bound, then we can conclude that a reduction in noise rate is necessarily needed for a possible quantum advantage. An approach to assessing the impact of noise on quantum algorithms is to directly simulate the circuit more so since the presence of noise in quantum circuits is expected to make them easier to classically simulate  \cite{aharonovPolynomialTimeClassicalAlgorithm2023, trivediTransitionsComputationalComplexity2022}. In fact, there have been several recent demonstrations of noisy quantum circuit simulations using tensor network methods \cite{zhouWhatLimitsSimulation2020, panSimulationQuantumCircuits2022, rakovszkyDissipationassistedOperatorEvolution2022}. However, most of the tensor network methods lack rigorous accuracy guarantees and cannot certify an accurate simulation of the quantum circuit. In particular, they are expected to deviate significantly from the circuit output as the noise rate continues to decrease and thus fall short of rigorously providing a no-go result for quantum advantage.

Alternatively, this problem can be approached analytically using tools from quantum information theory. For instance, Refs.~\cite{aharonovLimitationsNoisyReversible1996a, stilckfrancaLimitationsOptimizationAlgorithms2021, de2023limitations} analyzed  the increase in entropy of the quantum state due to noise, and showed that it can allow for an analytical lower bound on the attainable minimum energy. However, while providing rigorous no-go results, these analyses were circuit-architecture independent and were thus expected to underestimate the impact of noise. Certain circuit architectures are expected to significantly worsen the impact of noise, and this phenomena has been theoretically demonstrated in random quantum circuits models \cite{gonzalez-garciaErrorPropagationNISQ2022, deshpandeTightBoundsConvergence2022}. However, it remains unclear if it is possible to provide an architecture-dependent lower bound for a specific engineered quantum circuit.

In this article, we propose a method for efficiently computing rigorous bounds on the performance of any \textit{specified} quantum circuit in the presence of a constant rate of depolarizing noise. The key insight behind our proposed method is the formulation of a Lagrangian dual corresponding to the circuit dynamics, which allows us to account for the circuit architecture in addition to the increase in the entropy, or equivalently, the decrease in the purity of the quantum state. We show that the Lagrangian dual yields a hierarchy of classically computable lower bounds on energy, with respect to a specified Hamiltonian, obtained at the output of a noisy quantum circuit. We provide numerical and analytical evidence that this formulation can capture the circuit-architecture dependent propagation of errors through the noisy quantum circuit and thus provide more stringent lower bounds than currently available. Our work is, in part, motivated by the application of Lagrangian duality to provide performance bounds on classical physical systems \cite{trivediBoundsScatteringAbsorptionless2020a, angeris2019computational, schabTradeoffsAbsorptionScattering2020, chaoPhysicalLimitsElectromagnetism2022, moleskyGlobalMathbbTOperator2020, moleskyHierarchicalMeanfieldMathbbT2020} and quantum optical devices \cite{zhangConservationLawBasedGlobalBounds2021, mishraControlDesignInhomogeneousBroadening2021}.
\newpage
\section{Notation}
Given a finite-dimensional Hilbert space $\mathcal{H}$, we use $\mathcal{D}_1(\mathcal{H})$ to denote the set of all density matrices on $\mathcal{H}$, and $\mathcal{M}(\mathcal{H})$ to denote the set of Hermitian linear operators on $\mathcal{H}$. Unless otherwise mentioned, for any linear operator $A$ on $\mathcal{H}$, $\norm{A}$ will denote its operator norm i.e.~the maximum singular value of $A$ and $\norm{A}_F = \sqrt{\text{Tr}(A^\dagger A)}$ will denote its Frobenius norm.

We use the computer-science big-O notation for function asymptotics. In particular, given two functions $f, g:[0, \infty) \to [0, \infty)$, $f = O(g)$ if, for some $c > 0$, $f(x) \leq c g(x)$ as $x \to \infty$ and $f = \Omega(g)$ if, for some $c > 0$, $f(x) \geq c g(x)$ as $x \to \infty$.

\section{Duality based bounds}\label{duality-based-bounds}
\subsection{Single-qubit example}
\begin{figure}[t]
    \centering
    \includegraphics[width=0.48\textwidth]{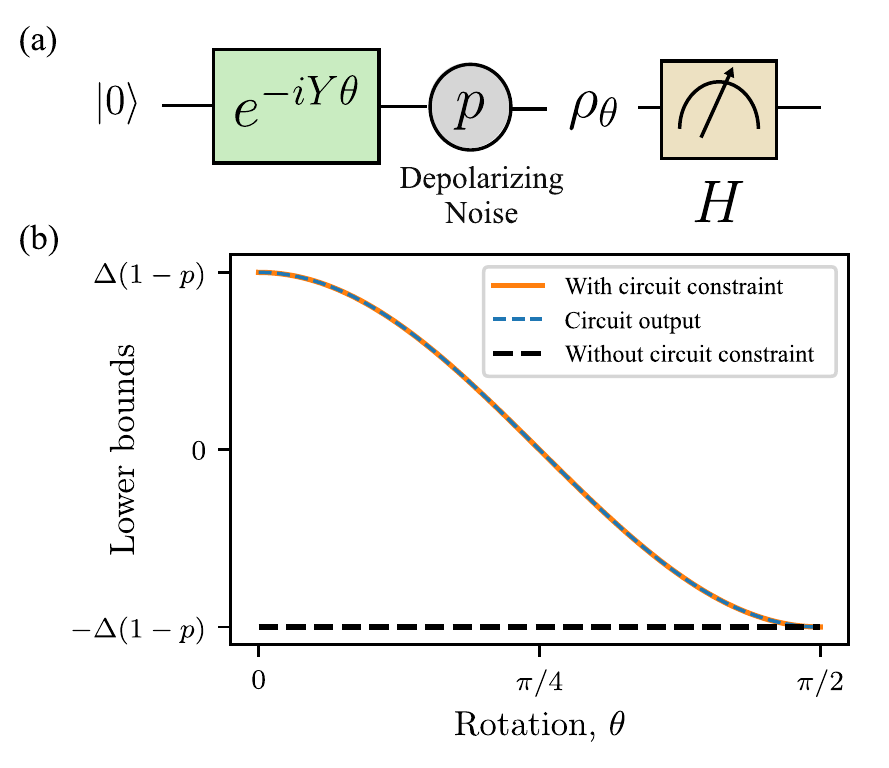}
    \caption{ Comparison of bounds, with and without accounting for circuit constraints, on the minimum energy corresponding to the Hamiltonian $H = \Delta Z$  attainable by the output of the single qubit circuit shown in the schematic. The circuit consists of a Y-axis rotation followed by depolarizing noise acting with probability $p$.} 
    \label{fig:sq}
\end{figure}

As a simple illustrative example of the Lagrangian dual formulation, we first consider a single-qubit circuit [Fig.~\ref{fig:sq}(a)] --- consider a qubit initially in $\ket{0}$, with a gate $U = e^{-i\theta Y}$ being applied on it followed by depolarizing noise with probability $p$. We would like to find the parameter $\theta$ to minimize the energy corresponding to the Hamiltonian $H = \Delta Z$ --- in the absence of noise ($p = 0$), it is straighforward to verify that this would be accomplished by setting $\theta = \pi/2$ to obtain an energy $-\Delta$.

However, in the presence of depolarizing noise, the qubit will necessarily be in a mixed state. The extent to which the state is mixed can be quantified with a purity measure, for instance the von Neumann entropy of the qubit state, its trace purity or even higher order Renyi Entropies \cite{muller2013quantum}. For concreteness, we use the trace purity measure of a state $\rho$: $P(\rho) = \text{Tr}(\rho^2)$ --- $P(\rho) = 1$ if and only if $\rho$ is a pure state, else $P(\rho) < 1$. Now, since the state $\rho_\theta$ at the output of the single-qubit circuit in Fig.~\ref{fig:sq}(a) is obtained by applying the depolarizing noise channel to a single qubit pure state, $P(\rho_\theta) = P_0 := p^2 / 4 + (1 - p/2)^2 < 1$. Since $\rho_\theta$ is necessarily mixed, it cannot produce the pure ground state of the Hamiltonian $H$ perfectly irrespective of the choice of $\theta$ -- in fact, this simple observation can be used to lower bound the energy that can possibly be obtained at the output of the circuit by minimizing it with respect to states with purity at-most $P_0$ i.e.~solving the following optimization problem
\begin{align}\label{eq:trace_purity_qubit_nockt}
        \underset{{\rho \in \mathcal{D}_1(\mathbb{C}^2)}}{\text{minimize}} \quad & \text{Tr}(H\rho)\nonumber\\
    \textrm{subject to} \quad &  P(\rho) \leq P_0,
\end{align}
where $\mathcal{D}_1(\mathbb{C}^d)$ is the space of density matrices on the Hilbert space $\mathbb{C}^d$. The optimization problem in Eq.~\ref{eq:trace_purity_qubit_nockt} is solved by $\rho = (1 - p/2) \ket{1}\bra{1} + (p/2) \ket{0}\bra{0}$ with energy $-\Delta(1 - p)$. This bound clearly exhibits the intuitively expected dependence on the noise rate $p$ --- if $p = 0$, then the energy attained coincides with the ground state energy of $-\Delta$, and if $p = 1$, it is simply the energy obtained by the maximally mixed state. 

However, this bound does not account for the unitary being applied on the qubit, and a better bound can be obtained by explicitly accounting for the circuit. To do so, we use the method of Lagrange duality \cite{boydConvexOptimization2004a, trivediBoundsScatteringAbsorptionless2020a}. For this, we extend the problem in Eq.~\ref{eq:trace_purity_qubit_nockt} by adding an additional constraint due to the circuit:
\begin{align*} \label{eq:primal}
    \underset{\rho \in \mathcal{D}_1(\mathbb{C}^2)}{\text{minimize}} \quad & \text{Tr}(H\rho)\\
    \textrm{subject to} \quad & \rho = \mathcal{E}_\theta(\rho_{0}),\\
    & P(\rho) \leq P_0, \numberthis
\end{align*}
where $\rho_0 = \ket{0}\bra{0}$ and $\mathcal{E}_\theta$ is the channel corresponding to the unitary $e^{-iY\theta}$ followed by the single-qubit depolarizing noise. Note that the purity constraint $P(\rho) \leq P_0$ is redundant and is already implied by the circuit constraint $\rho = \mathcal{E}_\theta(\rho_0)$. However, as we will see in the following discussion, while redundant constraints do not impact the solution of an optimization problem, depending on the specific technique used to obtain a lower bound on the problem, they can have a considerable impact.

To provide a lower bound on this optimization, we construct its Lagrangian $\mathcal{L}(\sigma, \lambda)$ by introducing Lagrange multipliers $\sigma \in\mathcal{M}( \mathbb{C}^2)$, and $\lambda \geq 0$,
\begin{align}\label{eq:lag_sq}
&\mathcal{L}(\rho, \sigma, \lambda)= \text{Tr}[H\rho] + \text{Tr}[\sigma(\rho - \mathcal{E}_\theta(\ket{0}\bra{0}))]\nonumber\\
&\qquad  + \lambda(P(\rho) - P_0).
\end{align}
$\mathcal{L}(\rho, \sigma, \lambda)$ can be considered to be a modified energy function which, in addition to the energy $\text{Tr}[H\rho]$, also penalizes violation of the two constraints: $\rho = \mathcal{E}_\theta(\rho_0)$ imposed by the circuit \emph{and} $P(\rho)  \leq P_0$ on the purity of the state $\rho$. Minimizing the Lagrangian with respect to $\rho$, we obtain the \emph{dual function},
\begin{align}\label{eq:dual_def}
g(\sigma, \lambda) = \min_\rho  \mathcal{L}(\rho, \sigma, \lambda),
\end{align}
which is a function of $\sigma, \lambda$, the \emph{dual variables}. It follows from the principle of Lagrange duality that for any $\sigma$ and $\lambda \geq 0$, $g(\sigma, \lambda)$ is a lower bound on the energy attained by the circuit. This can easily be seen from Eq.~\ref{eq:lag_sq} by noting that when $\mathcal{L}$ is evaluated at the circuit output $\rho_\theta = \mathcal{E}_\theta(\ket{0}\bra{0})$, we obtain 
\[
\mathcal{L}(\rho_\theta, \sigma, \lambda) = \text{Tr}[H\rho_\theta] + \lambda(P(\rho_\theta) - P_0) \leq \text{Tr}[H\rho_\theta],
\]
since $P(\rho_\theta) \leq P_0$ and $\lambda \geq 0$. Since from Eq.~\ref{eq:dual_def} $g(\sigma, \lambda)$ is the smallest attainable value of $\mathcal{L}(\rho, \sigma, \lambda)$ on varying $\rho$, we obtain $g(\sigma, \lambda) \leq \text{Tr}[H\rho_\theta]$. We emphasize that the dual function $g(\sigma, \lambda)$, evaluated at any $\sigma, \lambda \geq 0$, is a lower bound on the energy $\text{Tr}[H\rho_\theta]$ attained by the circuit, and the best lower bound can be obtained by maximizing $g(\sigma, \lambda)$ with respect to $\sigma, \lambda$. Furthermore, since the construction of the dual function explicitly accounts for the circuit constraint, it gives a better bound than obtained from the problem in Eq.~\ref{eq:trace_purity_qubit_nockt} i.e.~by just accounting for the final purity of the state. This is exhibited in Fig.~\ref{fig:sq}(b), where $\text{max}_{\sigma, \lambda \geq 0}g(\sigma, \lambda)$ compared with $-\Delta(1 - p)$ and it can be seen that the dual function provides a better lower bound for most values of $\theta$. For the simple example of a single qubit, the duality-based bound that we can compute coincides exactly with the circuit output and thus models it exactly. As we will see in the next sections, this will not be the case for circuits over a large number of qubits.

\begin{figure}[t]
    \centering
    \includegraphics[width=0.5\textwidth]{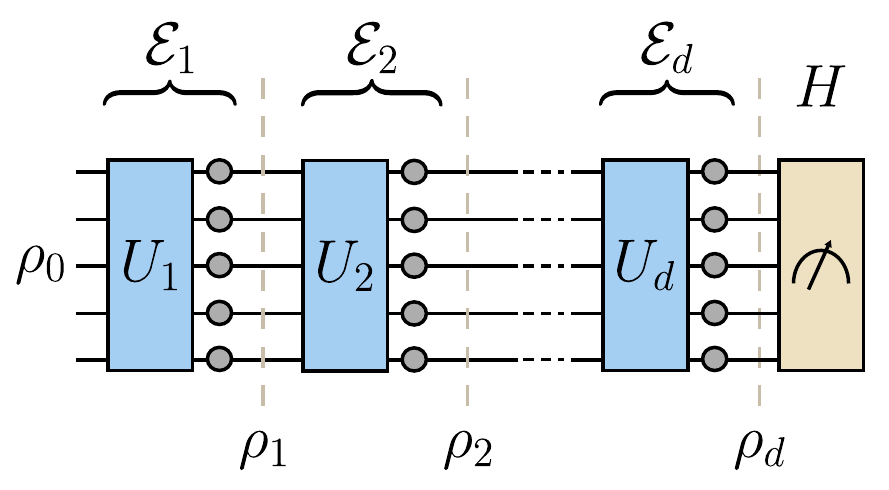}
    \caption{Schematic depiction of the problem setting considered in this paper. The unitaries (colored boxes) implement a quantum algorithm to prepare an approximation of the ground state of a target Hamiltonian $H$ in the absence of noise. Each layer of unitary is followed by single-qubit depolarizing noise (gray circles) on the qubits applied with a probability $p$.} 
    \label{fig:convention}
\end{figure}

\subsection{General formulation}
We can now extend the duality lower bound to more general quantum circuits [Fig. \ref{fig:convention}] --- consider a quantum circuit of depth $d$, consisting of unitaries $U_1, U_2 \dots U_d$ that has been designed to approximate the ground state of a target Hamiltonian $H$ of $N$ qubits. In the presence of noise, the state becomes increasingly mixed as the unitaries are applied on it --- while it is typically hard to compute exactly how mixed the state is, an analytical upper bound on several purity measures can be obtained. In particular, Refs.~\cite{aharonovLimitationsNoisyReversible1996a, de2023limitations} establish explicit upper bounds for two purity measures after $t$ time-steps --- the information content of the quantum state, as well as its trace purity.
\begin{lemma}\label{lemma:depolarizing_entropy_bounds}[Refs.~\cite{aharonovLimitationsNoisyReversible1996a}, \cite{de2023limitations}]
Suppose $\rho_t$ is the quantum state on $N$ qubits obtained from an initial pure state after applying $t$ unitaries and single qubit depolarizing channels, then
\begin{align*}
&I(\rho_t) := N + \textnormal{Tr}[\rho_t \log_2(\rho_t)] \leq N (1 - p)^t, \\
&P_\textnormal{tr}(\rho_t) := \textnormal{Tr}(\rho_t^2) \leq 2^{-N(1 - (1 - p)^t)},
\end{align*}
where $p$ is the probability of applying the depolarizing noise at each time-step independently on each qubit.
\end{lemma}
 \noindent Both the information content and trace purity can be viewed as measures of how mixed the given state is. Both are largest for a pure state ($I(\ket{\psi}\!\bra{\psi}) = N$ and $P_\text{tr}(\ket{\psi}\!\bra{\psi}) = 1$), and are lowest for the maximally mixed state ($I(I/2^N) = 0$ and $P_\textnormal{tr}(I/2^N) = 2^{-N}$).

In the remainder of this subsection, we denote by $P_t$ an upper bound on the purity of the state at $t^\text{th}$ time-step --- we will formulate the results of this subsection for general convex purity measures, and specialize them to concrete purity measures (such as information content or trace purity) in the following sections. Now, as with the single-qubit case, the energy attained at the output of the circuit can now be written as,
\begin{align} \label{eq:n_qubit_primal_problem}
    \underset{{\rho_1, \rho_2 \dots \rho_d} \in \mathcal{S}}{\text{minimize}} \quad & \text{Tr}(H\rho_d) \nonumber\\
    \textrm{subject to} \quad & \rho_t = \mathcal{E}_t(\rho_{t-1}), \ t \in \{1, \ldots, d \},\nonumber \\
    & P(\rho_t) \leq P_t, \ t \in \{1, \ldots, d \},\numberthis
\end{align}
where $\mathcal{E}_t(\cdot)$ is the quantum channel that applies the unitary $U_t$ for the $t^\text{th}$ layer of the circuit followed by depolarizing noise  acting individually on the qubits, and $\rho_0$ is a fixed and known initial state. Importantly, in Eq.~\ref{eq:n_qubit_primal_problem}, $\mathcal{S}$ is the set of $N$ qubit operators over which we allow the states $\rho_1, \rho_2 \dots \rho_d$ to vary --- this set can be chosen to be \emph{any} set containing density matrices over $N$ qubits $\mathcal{D}_1((\mathbb{C}^2)^{\otimes N})$ since the circuit constraints ($\rho_t = \mathcal{E}_t(\rho_{t-1})$) enforce $\rho_1, \rho_2 \dots \rho_d$ to be valid density matrices. For instance, $\mathcal{S}$ can be chosen to be just the set of $N$-qubit Hermitian operators, or the set of $N$-qubit Hermitian operators with unity trace. As we will see below, the choice of this set together with the purity measure determines the form of the dual function. 

To construct the dual function corresponding to Eq.~\ref{eq:n_qubit_primal_problem} --- we introduce the dual variables $\vec{\sigma} = \{\sigma_1, \sigma_2 \dots \sigma_d\}$, which are $N-$qubit Hermitian operators, corresponding to the circuit constraints and $\vec{\lambda} = \{\lambda_1, \lambda_2 \dots \lambda_d \geq 0\}$ corresponding to the purity constraints. The Lagrangian is now constructed by adding penalties corresponding to the circuit constraints and purity constraints at each time-step to the output energy:
\label{eq:lag}
\begin{align}
&\mathcal{L}(\vec{\rho}, \vec{\sigma}, \vec{\lambda}) = \text{Tr}[H\rho_d] + \sum_{t = 1}^d \text{Tr}\bigg[\sigma_t\big(\rho_t - \mathcal{E}_t(\rho_{t - 1})\big) \bigg] + \nonumber \\
&\qquad \qquad \qquad \qquad \sum_{t = 1}^d \lambda_t\bigg[P(\rho_t) - P_t \bigg], \nonumber \\
&\qquad \qquad = \sum_{t = 1}^d \text{Tr}{(\rho_t H_t)} + \lambda_t \bigg[P(\rho_t) - P_t\bigg],
\end{align}
where $H_d = H + \sigma_d$, $H_t = \sigma_t - \mathcal{E}_{t + 1}^\dagger(\sigma_{t + 1})$ for $t \in \{1, 2 \dots d-1\}$ 
The dual function is obtained by minimizing the Lagrangian with respect to $\rho_1, \rho_2 \dots \rho_d \in \mathcal{S}$ i.e.
\begin{subequations}\label{eq:dual_function}
    \begin{align}
    g(\vec{\sigma}, \vec{\lambda}) &= \min_{\rho_1, \rho_2 \dots \rho_d \in \mathcal{S}} \mathcal{L}(\vec{\rho}, \vec{\sigma}, \vec{\lambda}), \nonumber \\
    &=\sum_{t = 1}^d \min_{\rho_t \in \mathcal{S}} \bigg[\big(\textnormal{Tr}(\rho_t H_t) + \lambda_t P(\rho_t)\big) - \lambda_t P_t\bigg], \nonumber \\
    &=\sum_{t = 1}^d \bigg(\mathcal{F}_{\mathcal{S}, P}(H_t, \lambda_t) - \lambda_t P_t\bigg)
    \end{align}
where
\begin{align} \label{eq:generalized_free_energy}
    \mathcal{F}_{\mathcal{S}, P}(H, \lambda) = \min_{\rho \in \mathcal{S}}\bigg( \textnormal{Tr}[H\rho] + \lambda P(\rho)\bigg). 
\end{align}
\end{subequations}
As with the single-qubit example, the dual function is a lower bound on the energy produced at the circuit output for any $\vec{\sigma}, \vec{\lambda} \geq 0$ i.e.
\[
g(\vec{\sigma}, \vec{\lambda}) \leq \textnormal{Tr}[\rho_d H] \ \textnormal{ for all }\sigma_t \in \mathcal{M}((\mathbb{C}^2)^{\otimes N}), \lambda_t \geq 0.
\]
The function $\mathcal{F}_{\mathcal{S}, P}(H, \lambda)$ can be interpreted as a generalized free energy corresponding to the Hamiltonian $H$ at temperature $\lambda$ which depends on both the domain $\mathcal{S}$ and the purity measure $P$. For instance, if the purity measure is taken to be the information content $I(\rho) = N - \textnormal{Tr}[\rho \log_2(\rho)]$, then it reduces to the Gibbs free energy with an offset of $N\lambda$. However, by choosing different purity measures $P$ as well as different domains $\mathcal{S}$, the dual function allows us to obtain a family of bounds on the noisy quantum circuit. As we will see in the next section, certain choices of $P$ and $\mathcal{S}$ provide lower bounds that can be classically computed.

Consider first the best lower bound that can be obtained from the dual function. In the following proposition, we show that the best lower bound attained by the dual function is exactly equal to the energy attained by the quantum circuit, and choice of dual variables $\sigma_1, \sigma_2 \dots \sigma_d$ that yields the largest value dual function corresponds to the Heisenberg picture evolution of the Hamiltonian $H$.
\begin{proposition}\label{prop:heisenberg_picture_exact_dual}
For the dual function defined in Eq.~\ref{eq:dual_function}, it follows that its maximum over the dual variables is equal to the output energy of the noisy circuit i.e.~
\[
\underset{\substack{\sigma_1, \sigma_2 \dots \sigma_d \in \mathcal{M}((\mathbb{C}^2)^{\otimes N}) \\
\lambda_1, \lambda_2 \dots \lambda_d \geq 0}}{\textnormal{maximum}} g(\vec{\sigma}, \vec{\lambda}) = \textnormal{Tr}[H\mathcal{E}_d \mathcal{E}_{d-1}\dots \mathcal{E}_1(\rho_0)],
\]
and the maximum is attained at
\[
\sigma_d = -H, \sigma_t = -\mathcal{E}_{t + 1}^\dagger \mathcal{E}_{t + 2}^\dagger \dots \mathcal{E}_d^\dagger(H),
\]
and $\lambda_1 = \lambda_2= \dots \lambda_d = 0$.
\end{proposition}
\noindent\emph{Proof}: The proof of this proposition follows simply by noting that, from definition,
\[
\mathcal{F}_{\mathcal{S}, P}(0, 0) = 0.
\]
Now, if $\sigma_d = -H$, and $\sigma_t = -\mathcal{E}^\dagger_{t + 1}\mathcal{E}_{t+2}^\dagger \dots \mathcal{E}_d^\dagger(H)$, then $H_t = 0$. Hence, we obtain that at this value of $\vec{\sigma}$ and at $\vec{\lambda} = 0$, $g(\vec{\sigma}, \vec{\lambda}) = \textnormal{Tr}[\rho_0 \mathcal{E}_1^\dagger \mathcal{E}_2^\dagger \dots \mathcal{E}_d^\dagger(H)] = \textnormal{Tr}[H\mathcal{E}_d \mathcal{E}_{d-1}\dots \mathcal{E}_1(\rho_0)]$. Since $\textnormal{Tr}[H\mathcal{E}_d \mathcal{E}_{d-1}\dots \mathcal{E}_1(\rho_0)]$ is also an upper bound of $g(\vec{\sigma}, \vec{\lambda})$, the proposition follows. 
\ \hfill$\square$

This proposition establishes that finding the best dual bound is equivalent to exactly simulating the circuit, which we expect to be hard to do on classical computers. This hardness fundamentally stems from the fact that the dual variables $\sigma_1, \sigma_2 \dots \sigma_d$ are operators in an exponentially large space. However, since the dual function $g(\vec{\sigma}, \vec{\lambda})$ is a lower bound on the output energy for any $\vec{\sigma}, \vec{\lambda}$, a natural approach to evaluate a lower bound would be restrict $\sigma_i$ to subsets of $\mathcal{M}((\mathbb{C}^2)^{\otimes N})$ where the dual function could be evaluated efficiently --- the specific subset would depend on the choice of the purity measure. In Section \ref{sec:purity}, we will see that the dual function obtained on choosing the purity measure to be trace purity and the domain $ \mathcal{S} = \mathcal{M}((\mathbb{C}^2)^{\otimes N})$ of $N$-qubit Hermitian operators can be evaluated efficiently if $\sigma_1, \sigma_2 \dots \sigma_d$ are chosen to be matrix product operators (MPOs) of bond-dimension $\text{poly}(N)$. In Section \ref{sec:information}, we will consider the dual function obtained on choosing the purity measure to be the information content of the state, in which case $\sigma_1, \sigma_2 \dots \sigma_d$ can be restricted to the space of geometrically local Hamiltonians, allowing for an exact evaluation of the dual function. 

Restricting the dual variables $\sigma_1, \sigma_2 \dots \sigma_d$ to a subset of  $\mathcal{M}((\mathbb{C}^2)^{\otimes N})$ raises the question of whether the maximum value that the dual function can attain within this restricted set of dual variables gives a better lower bound on the energy compared to neglecting the circuit constraints and just accounting for the purity of the final state i.e.~does the duality based bound still account for the circuit architecture. Our next proposition answers this question affirmatively, and shows that a better lower bound can be obtained as long as the restricted set of dual variables contains $0$.

\begin{proposition}
    Suppose $\mathcal{S}_\sigma \subset \mathcal{M}((\mathbb{C}^2)^{\otimes N})$, such that $0 \in \mathcal{S}_\sigma$, then 
    \[
    \underset{\substack{\sigma_1, \sigma_2 \dots \sigma_d \in \mathcal{S}_\sigma \\
\lambda_1, \lambda_2 \dots \lambda_d \geq 0}}{\textnormal{maximum}}\ g(\vec{\sigma}, \vec{\lambda}) \geq    \underset{\rho \in \mathcal{S}, P(\rho) \leq P_d}{\textnormal{minimize}} \  \textnormal{Tr}(H\rho)
    \]
\end{proposition}
\noindent\emph{Proof}: Since $0 \in \mathcal{S}_\sigma$,
\[
\underset{\lambda \geq 0}{\textnormal{maximum}}\ g(\{0 \dots 0\}, \{0\dots \lambda\}) \leq \underset{\substack{\sigma_1, \sigma_2 \dots \sigma_d \in \mathcal{S}_\sigma \\
\lambda_1, \lambda_2 \dots \lambda_d \geq 0}}{\textnormal{maximum}}\ g(\vec{\sigma}, \vec{\lambda}).
\]
Now, we can note that
\[
g(\{0 \dots 0\}, \{0\dots \lambda\})= \mathcal{F}_{S, P}(H, \lambda) - \lambda P_d.
\]
It can be noted that $g(\{0 \dots 0\}, \{0 \dots \lambda\})$ is simply the dual function of the convex problem
\begin{align*}
        \underset{{\rho \in \mathcal{S}}}{\textnormal{minimize}} \quad & \text{Tr}(H\rho)\nonumber\\
    \textrm{subject to} \quad &  P(\rho) \leq P_d.
\end{align*}
Furthermore, this convex problem trivially satisfies the Slater's conditions \cite{slaterLagrangeMultipliersRevisited2014, boydConvexOptimization2004a}. This can be checked by noting that the Slater's conditions are satisfied if there is a $\rho \in \mathcal{S}$ such that $P(\rho) < P_d$ --- this follows by noting that $P(I/2^N) < P_d$ and $I/2^N \in \mathcal{D}_1((\mathbb{C}^2)^{\otimes N}) \subseteq \mathcal{S}$. Since Slater's conditions are satisfied, this problem is strongly dual and consequently the optimal duality-bound is equal to the solution of the optimization problem i.e.
\[
\underset{\lambda \geq 0}{\textnormal{maximum}}\ g(\{0 \dots 0\}, \{0\dots \lambda\}) = \underset{\rho \in \mathcal{S}, P(\rho) \leq P_d}{\textnormal{minimize}} \  \textnormal{Tr}(H\rho),
\]
which proves the proposition. \hfill $\square$

While this proposition indicates that accounting for the circuit constraints while constructing the lower bound results in an improvement over only accounting for the final purity even with restricted space of dual variables, it says nothing about the extent to which the bound improves. We expect the improvement to be strongly dependent on the purity function $P$, the domain set $\mathcal{S}$, and the dual set $\mathcal{S}_\sigma$ used in formulating and evaluating the bound. In the next section, we consider a specific formulation of the dual function that uses the trace purity measure, and show that the lower bound obtained on accounting for the circuit constraints can be exponentially better than if the circuit constraints were not accounted for.

\section{Trace purity-based lower bound} \label{sec:purity}
\subsection{Formulation}
In this section, we consider now a specific choice of the purity function and the domain $\mathcal{S}$ that results in a dual function that can be computed exactly when the dual variables are parametrized as matrix product operators with $\textnormal{poly}(N)$ bond dimension. We choose the purity measure to be trace purity $P(\rho) = P_\text{tr}(\rho) = \textnormal{Tr}(\rho^2)$, and the domain $\mathcal{S}$ in Eq.~\ref{eq:n_qubit_primal_problem} to be the space of Hermitian $N$-qubit operators $\mathcal{M}((\mathbb{C}^2)^{\otimes N})$. It then follows that $\mathcal{F}_{\mathcal{S}, P}(H, \lambda)$ defined in Eq.~\ref{eq:generalized_free_energy} evaluates to
\[
\mathcal{F}_{\mathcal{S}, P}(H, \lambda) = -\frac{\textnormal{Tr}(H^2)}{4\lambda},
\]
and therefore, we obtain that
\[
g(\vec{\sigma}, \vec{\lambda}) = -\textnormal{Tr}[\rho_0 \mathcal{E}^\dagger_1(\sigma_1)] - \sum_{t = 1}^d \bigg(\frac{\textnormal{Tr}(H_t^2)}{4\lambda_t}+ \lambda_t P_t\bigg),
\]
where $H_d = H + \sigma_d $ and $H_t = \sigma_t - \mathcal{E}_{t + 1}^\dagger (\sigma_{t + 1})$. Furthermore, for this dual function, it is possible to perform the maximization over $\vec{\lambda}$ analytically to obtain
\begin{align}\label{eq:duality_bound_trace_purity}
h(\vec{\sigma}) &= \underset{{\vec{\lambda} \geq 0}}{\text{maximum}} \quad g(\vec{\sigma}, \vec{\lambda}), \nonumber\\
&= -\textnormal{Tr}[\rho_0 \mathcal{E}_1^\dagger(\sigma_1)] - \sum_{i = 1}^d \sqrt{P_t \textnormal{Tr}(H_t^2)}.
\end{align}
From the expression for $h(\vec{\sigma})$, we immediately notice that if $\sigma_1, \sigma_2 \dots \sigma_d$ are restricted to be matrix product operators with bond dimension $D$, then $h(\vec{\sigma})$ can be evaluated classically in time $NdD^4$. However, as we established in proposition 1, the best lower bound is obtained $h(\vec{\sigma})$ when evaluating it at $\vec{\sigma}$ corresponding to a Heisenberg picture evolution of Hamiltonian $H$. While for most problems of interest (e.g.~where $H$ is a local or spatially-local Hamiltonian), $H$ can be represented as a matrix product operator of a modest bond dimension, the unitaries involved in the circuit can, in general, grow its bond dimension exponentially. A natural choice of $\sigma_1, \sigma_2 \dots \sigma_d$ would then be to perform time-evolving block decimation (TEBD) \cite{vidalEfficientSimulationOneDimensional2004, verstraeteMatrixProductDensity2004a, daleyTimedependentDensitymatrixRenormalizationgroup2004, whiteRealTimeEvolutionUsing2004, vidalEfficientClassicalSimulation2003} on the Heisenberg evolution and compress the operators in each step into bond-dimension $D$ i.e. at $\vec{\sigma}^{h, D} = \{\sigma_1^{h, D}, \sigma_2^{h, D},\dots,  \sigma_d^{h, D}\}$
\begin{align}\label{eq:heis_tebd}
&\sigma_d^{h, D} = -H \text{ and, }\nonumber \\
&\sigma_{t}^{h, D} = \Pi_D \mathcal{E}_{t + 1}^\dagger(\sigma_{t + 1}^{h, D}) \text{ for }t\in\{1, 2 \dots d -1\},
\end{align}
where $\Pi_D$ compresses an $N$-qubit operator to an operator with a bond-dimension $D$ \cite{schollwock2011density}.

\emph{Duality-bound and TEBD truncation errors}. If a Heisenberg picture TEBD simulation, for some bond dimension $D$, of the noisy quantum circuit is exact, then by Proposition~\ref{prop:heisenberg_picture_exact_dual}, the duality based bound $h(\vec{\sigma}^{h, D})$ is exactly equal to the expected energy at the output of the circuit. In practice, for small bond dimensions $D$, the TEBD algorithm is not exactly correct but incurs an error. However, as shown below, an upper bound on this error can also be efficiently computed for the TEBD algorithm. Consequently, tracking the error incurred in the TEBD algorithm allows us to calculate another lower bound on the output of the quantum circuit i.e.~if the TEBD algorithm produces an estimate $E_\text{TEBD}$ the output energy $E$ of the circuit within an additive error $\delta$, then $E_\text{TEBD} - \delta$ also lower bounds the energy $E$. A natural question to ask is if the duality based bounds are more informative than the bound obtained from just a TEBD simulation.

Consider now the problem of estimating the TEBD error following Ref.~\cite{verstraeteMatrixProductDensity2004a}. The TEBD estimate of the energy at the circuit output, $E_\text{TEBD}$, can be expressed as
\[
E_\text{TEBD} = \textnormal{Tr}\bigg(\mathcal{E}_1(\rho_0) \bigg(\prod_{t = 2}^d \Pi_D \mathcal{E}_t^\dagger\bigg) (H)\bigg),
\]
while the true energy at the circuit output can be expressed as
\[
E =  \textnormal{Tr}\bigg(\mathcal{E}_1(\rho_0) \bigg(\prod_{t = 2}^d \mathcal{E}_t^\dagger\bigg) (H)\bigg).
\]
Denoting by $\rho_t$ the state of the qubits in the quantum circuit at time-step $t$, $\rho_t = \mathcal{E}_t \mathcal{E}_{t - 1} \dots \mathcal{E}_1(\rho_0)$, we note that
\begin{align*}
    &E - E_\text{TEBD} \nonumber\\
    &= \textnormal{Tr}\bigg(\rho_1 \bigg(\prod_{t = 2}^d \mathcal{E}_t^\dagger - \prod_{t = 2}^d \Pi_D \mathcal{E}_t^\dagger\bigg)(H)\bigg), \nonumber \\
    &=\sum_{t =2}^d \textnormal{Tr}\bigg(\rho_1 \bigg(\prod_{s = 2}^{t - 1}\mathcal{E}_s^\dagger \bigg)\bigg(\mathcal{E}_t^\dagger - \Pi_D \mathcal{E}_t^\dagger\bigg)\bigg(\prod_{s = t + 1}^d \Pi_D \mathcal{E}_s^\dagger\bigg)(H)\bigg), \nonumber\\
    &=\sum_{t = 2}^d \text{Tr}\bigg(\rho_{t - 1} \bigg(\sigma_{t - 1}^{h, D} - \mathcal{E}_t^\dagger(\sigma_{t}^{h, D})\bigg)\bigg) =\sum_{t = 1}^{d - 1} \text{Tr}\big(\rho_{t} H_{t}\big),
\end{align*}
where, in the last step, we have used the fact that, by definition, $H_t = \sigma_{t - 1}^{h, D} - \mathcal{E}_t^\dagger(\sigma_t^{h, D}))$. Now, an upper bound on the error $\abs{E - E_\text{TEBD}}$ can be obtained via
\begin{align}\label{eq:tebd_tr_ineq}
    \abs{E - E_\text{TEBD}} \leq \sum_{t = 1}^{d - 1}\abs{\text{Tr}(\rho_t H_t)} \leq \sum_{t = 1}^{d-1}\norm{H_t}_F,
\end{align}
where $\norm{A}_F = \sqrt{\text{Tr}(A^\dagger A)}$ and we have used the fact that, by the Holder's inequality, $\abs{\text{Tr}(\rho_t H_t)} \leq \norm{\rho_t}_1 \norm{H_t} \leq \norm{H_t}_F$ since $\norm{\rho_t}_1 = 1$ and $\norm{H_t} \leq \norm{H_t}_F$. We point out that an important reason why we express the error bound in terms of the Frobenius norm of $H_t$, instead of its operator norm, is because the Frobenius norm can be efficiently computed if $H_t$ is a matrix product operator of a small bond dimension (which is the case while performing the TEBD simulation). The deviation bound in Eq.~\ref{eq:tebd_tr_ineq} implies a lower bound
\begin{align}\label{eq:bound_error_heis_ac}
E \geq E_\text{TEBD} - \delta = -\textnormal{Tr}[\rho_0 \mathcal{E}_1^\dagger(\sigma_1)] - \sum_{i = 1}^d \sqrt{\textnormal{Tr}(H_t^2)}.
\end{align}
This bound is significantly worse than the duality-based bound in Eq.~\ref{eq:duality_bound_trace_purity} as $P_t \ll 1$ for all time steps $t$. The key reason why just accounting for a worst-case accumulation of TEBD errors yields a loose lower bound is that the upper bound in $\abs{E - E_\textnormal{TEBD}}$ \emph{does not }account for the decrease in the trace purity of the quantum state in the presence of noise, which is explicitly factored into the formulation of the dual.
% In the following proposition, we show that $h(\vec{\sigma}^{h, D})$ can also be interpreted in terms of the TEBD compression errors, but with the added consideration of explicitly taking into account the purity decrease at each time step.
% \begin{proposition}
%     Suppose $H$ is a Hamiltonian which can be exactly represented as a MPO of bond dimension $D$, then
%     \[
%      \abs{E_\textnormal{TEBD} - E} \leq E_\textnormal{TEBD} - h(\vec{\sigma}^{h, D}) = \sum_{t=1}^d \sqrt{P_t \textnormal{Tr}(H_t^2)} 
%     \]
% \end{proposition}
% \noindent\emph{Proof}:  We have that, 
% \[
% E_\textnormal{TEBD} = \textnormal{Tr}\bigg(\mathcal{E}_1(\rho_0) \bigg(\prod_{t = 2}^d \Pi_D \mathcal{E}_t^\dagger\bigg) (H)\bigg),
% \]
% and
% \[
% E = \textnormal{Tr}\bigg(\mathcal{E}_1(\rho_0) \bigg(\prod_{t = 2}^d \mathcal{E}_t^\dagger\bigg) (H)\bigg).
% \]
% Defining $\rho_t = \mathcal{E}_t \mathcal{E}_{t - 1}\dots \mathcal{E}_1(\rho_0)$, we the obtain that
% \[
% E - E_\textnormal{TEBD} = \sum_{t = 1}^{d - 1}\textnormal{Tr}\big(\rho_t H_t \big),
% \]
% where $H_t = \sigma_t^{h, D} - \mathcal{E}_{t + 1}^\dagger(\sigma_{t + 1}^{h, D})$. It then follows that
% \begin{align*} 
% \abs{E - E_\textnormal{TEBD}} \leq \sum_{t = 1}^{d-1}\abs{\textnormal{Tr}(\rho_t H_t)} \leq \sum_{t = 1}^{d-1}\sqrt{\textnormal{Tr}(\rho_t^2) \textnormal{Tr}(H_t^2)}.
% \end{align*}
% Using $\textnormal{Tr}(\rho_t^2) \leq P_t$, we obtain the proposition. \hfill $\square$

\begin{figure}[t]
    \centering
    \includegraphics[width=0.45\textwidth]{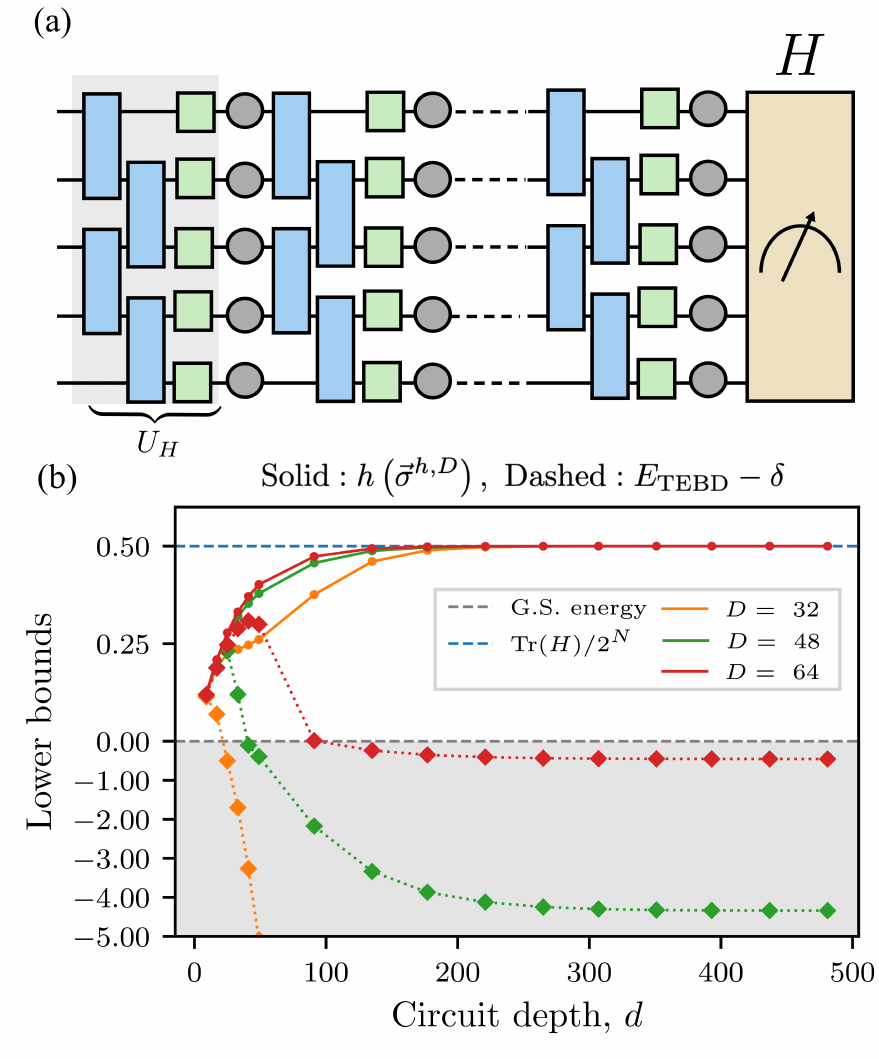}
    \caption{(a) Schematic of benchmark circuits considered for 1D spin systems: colored boxes indicate unitaries and grey circles depolarizing noise. Two-qubit unitaries are chosen to be $\exp(-i \theta X \otimes X)$ and single-qubit unitaries are independently Haar random. The Hamiltonian is chosen to be $H = -U_H \big(\sum_i Z_i \big) U_H^\dagger$, where $U_H$ is the first layer of unitaries, making $H$ a 4-local commuting Hamiltonian. The first layer of unitaries $U_H$ thus transforms the initial state $\ket{0}^{\otimes N}$ into the ground state of $H$. The last $(d-1)/2$ layers are chosen to be the inverse of the previous $(d-1)/2$ layers --- in the absence of noise, the output of the circuit is the ground state of $H$. (b)  Plot shows trace purity-based dual bound ($h(\vec{\sigma}^{h, D})$ in Eq.~\ref{eq:duality_bound_trace_purity}) (solid lines, circular markers) and bound obtained by only considering the TEBD errors ($E_{\text{TEBD}} - \delta$ in Eq.~\ref{eq:bound_error_heis_ac}) (dotted lines, diamond markers) for the ground state (G. S.) energy of the target Hamiltonian, as a function of circuit depth $d$ for a system of $N = 40$ spins, with two-qubit gate parameter $\theta = 0.05$, depolarizing noise rate of $p = 3\%$ and varying MPO ansatz bond dimensions $D$. Grey dashed line indicates G.S. energy, grey shaded area indicates region of trivial bounds (less than G.S. energy), blue dashed line indicates energy of the completely mixed state $\mathds{1}/2^N$. The $y$-axis is scaled by a constant multiplicative factor in the trivial region for visual clarity. The Hamiltonian is shifted and scaled such that its spectrum is in $[0,1]$.}
    \label{fig:spinsheis}
\end{figure}

In Fig.~\ref{fig:spinsheis}, we numerically exhibit the difference between the bound in Eq.~\ref{eq:bound_error_heis_ac} and $h(\vec{\sigma}^{h, D})$ for a 1D circuit on $N = 40$ qubits [Fig.~\ref{fig:spinsheis}(a)], which is chosen to find the ground state of a commuting 1D Hamiltonian (see the figure caption for the exact circuit and Hamiltonian). As can be seen from Fig.~\ref{fig:spinsheis}(b), the lower bound computed from the trace purity based dual is significantly larger, and thus more representative of the impact of noise on the output energy, than the lower bound provided by Eq.~\ref{eq:bound_error_heis_ac}. {We point out that the dual variables $\vec{\sigma}^{h, D}$ obtained by TEBD in the Heisenberg picture are not necessarily the globally optimal choice in the space of all MPOs with bond dimension $D$ to evaluate the dual function $h(\vec{\sigma})$. The function $h(\vec{\sigma})$ can potentially be optimized beyond the TEBD-based value to obtain better lower bounds. In practice, we observe that local optimization of $h(\vec{\sigma})$ with a gradient-based method starting from the initial point of $\vec{\sigma} = \vec{\sigma}^{h,D}$ yields only a modest improvement over $h \left(\vec{\sigma}^{h, D}\right)$.}

\begin{figure*}[t]
    \centering
    \includegraphics[width=0.8\textwidth]{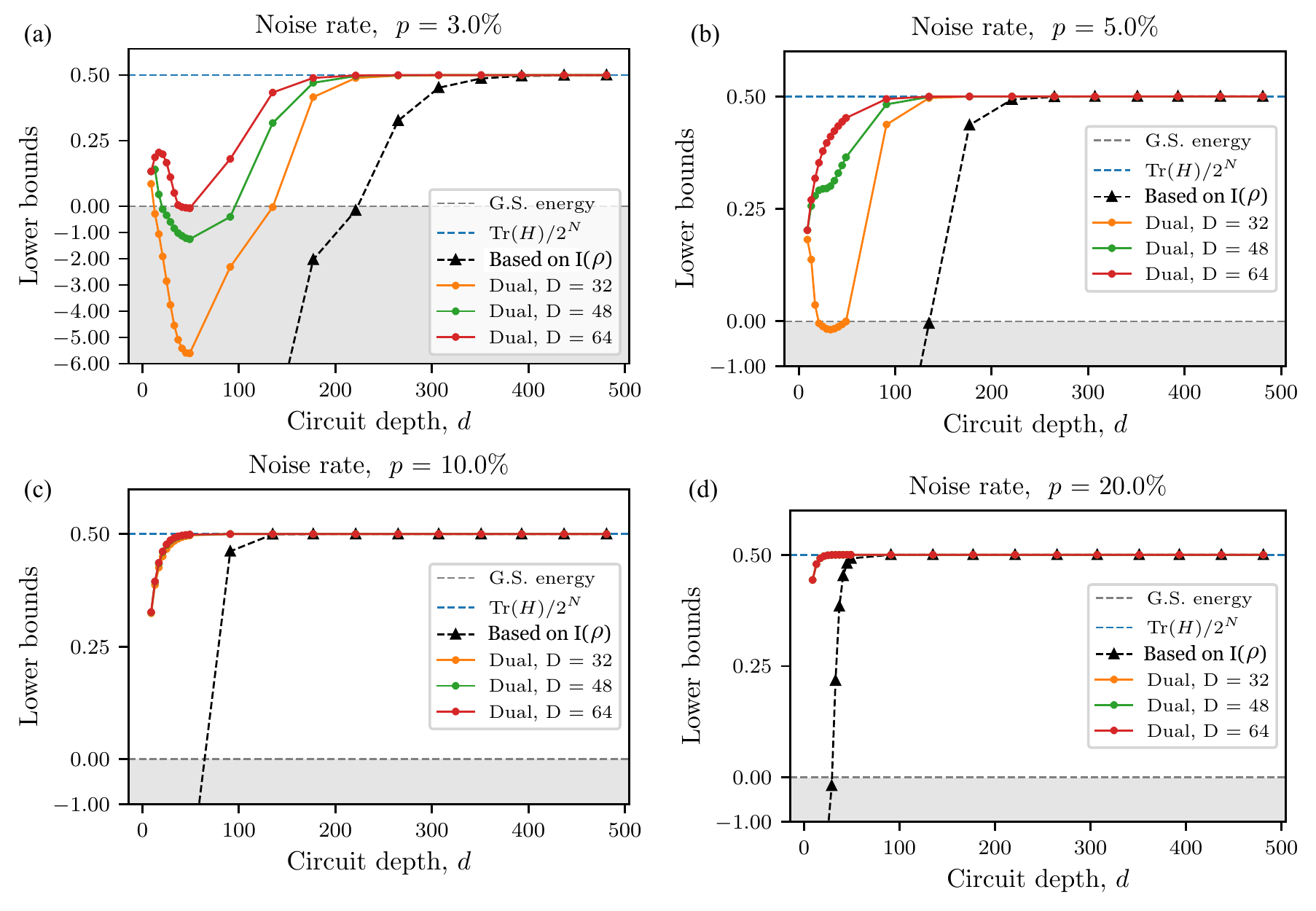}
    \caption{Comparison of the trace purity-based dual bound ($h(\vec{\sigma}^{h, D})$ in Eq.~\ref{eq:duality_bound_trace_purity}) and the bound based on just the information content of the output state ($\ell^I_{\lambda_c}$ in Eq.~\ref{eq:lbound_temp_ic}) for 1D many-body spin systems. Both bounds are lower bounds on the ground state (G. S.) energy of the same target Hamiltonian as considered for the results in Fig.~\ref{fig:spinsheis} (see description of Hamiltonian in caption of Fig.~\ref{fig:spinsheis}). Plots show the bounds as a function of brick-wall quantum circuit depth $d$ [Fig.~\ref{fig:spinsheis}(a)] for a system of $N = 40$ spins, varying MPO ansatz bond dimensions $D$, with depolarizing noise rates of (a) $p = 3\%$, (b) $p = 5\%$, (c) $p = 10 \%$, and (d) $p = 20\%$.  Note that at $p$ = 10\% and 20\%, approximately the same duality-based lower bound is obtained for different bond dimensions.} Two-qubit unitaries in the brick-wall circuit are chosen to be  $\exp(-i \theta X \otimes X)$ with $\theta = 0.1$ and single-qubit unitaries are independently Haar random. Grey dashed line indicates G.S. energy, grey shaded area indicates region of trivial bounds (less than G.S. energy), blue dashed line indicates energy of the completely mixed state $\mathds{1}/2^N$. The $y$-axis is scaled by a constant multiplicative factor in the trivial region for visual clarity. 
    The Hamiltonian is shifted and scaled such that its spectrum is in $[0,1]$.
    \label{fig:spinsentropic}
\end{figure*}

Next, we study the improvement that the duality based bounds that account for the circuit constraint provide over bounds in existing literature that just account for the information content at the circuit output. In particular, we numerically compare the best lower bound $\ell^\text{dual}_D$ that we can obtain by evaluating $h(\vec{\sigma})$ at $\sigma_1, \sigma_2 \dots \sigma_d \in \textnormal{MPO}_D$ (the space of all $N$-qubit MPOs of bond dimension $D$), 
\[
\ell^\text{dual}_D = \underset{\sigma_1, \sigma_2 \dots \sigma_d \in \textnormal{MPO}_D}{\text{maximize}} h(\vec{\sigma}),
\]
to the lower bound $\ell^I$ analyzed in Ref.~\cite{stilckfrancaLimitationsOptimizationAlgorithms2021},
\[
\ell^I = \underset{\rho: I(\rho) \leq N(1-p)^d}{\text{minimize}} \textnormal{Tr}(H\rho),
\]
i.e.~where they accounted only for the decreased information content $I(\rho) = N - \text{Tr}[\rho \log_2 \rho]$ of the final state as per Lemma 1. First, we show that there exists a Hamiltonian and a 1D circuit where $\ell^\text{dual}_D$, with $D = O(N)$ \cite{notation}, scales super-exponentially with the depth of the circuit and thus captures the propagation of errors through the circuit, while $\ell^I$ scales at-most exponentially with the circuit depth. 
\begin{proposition}
There exists a 1D circuit and a $N$-qubit Hamiltonian $H$ with $\textnormal{Tr}(H) = 0$ and $\norm{H} = N$, such that $\ell^{I} = -N(1-p)^{{O}(d)}$, while $ \ell^\textnormal{dual}_D = -N(1-p)^{\Omega(d^2)} + O(\sqrt{N})$ for a choice of $D \leq O(N)$.
\end{proposition}

\noindent\emph{Proof sketch (see appendix \ref{app:proof_prop_4} for details)}: Consider a Clifford circuit chosen at random from the ensemble of entangle-unentangle circuits analyzed in Ref.~\cite{gonzalez-garciaErrorPropagationNISQ2022} --- it was shown for this ensemble that, for a 1D circuit, on average, the energy of the output state with respect to a 2-local Hamiltonian converges to the energy of the maximally mixed state as  $\sim \text{poly}(N) \times (1-p)^{\Omega(d^2)}$. Consider now the Hamiltonian $H = -\sum_{i = 1}^{N} Z_i$ and initial state $\ket{0}^{\otimes N}$. In the Heisenberg picture, each $Z_i$ will be mapped to exactly one Pauli string under the action of Clifford gates \cite{gottesmanHeisenbergRepresentationQuantum1998a, aaronsonImprovedSimulationStabilizer2004a}. Since a Pauli string is representable as an MPO of bond dimension 1, $\sigma_t$ obtained from Heisenberg picture evolution will be a sum of $N$ MPOs of bond-dimension 1 and will thus be a MPO of bond-dimension at most $N$. Thus, from Proposition 1, the purity-based dual exactly matches the energy of the output of the quantum circuit, which will scale as $-N(1-p)^{\Omega(d^2)}$ \cite{gonzalez-garciaErrorPropagationNISQ2022, quekExponentiallyTighterBounds2023}. As the bound without circuit constraints is agnostic to the unitaries in a circuit, it is also a lower bound on the circuit where all the unitaries are just the identity operation. For this trivial circuit, the energy of the state after $d$ layers of just depolarizing noise scales as $-N(1-p)^d$. Hence, the lower bound without circuit constraints $\ell^I = -N(1-p)^{{O}(d)}.$ 

\subsection{Numerical studies}
For non-Clifford circuits, MPO parametrization with bond dimension $\text{poly}(N)$ is no longer expected to fully capture Heisenberg picture evolution. Even so, we numerically demonstrate for a 1D spin system that the circuit dual function Eq.~\ref{eq:duality_bound_trace_purity} can be used to compute circuit-specific bounds that perform better than bounds that only take into account the information content of the output state. We compute the bounds by evaluating the dual function $h(\vec{\sigma})$ at the dual variables obtained from TEBD on the Heisenberg picture as in Eq.~\ref{eq:heis_tebd}. Fig.~\ref{fig:spinsentropic} shows numerical studies of the bounds $h(\vec{\sigma}^{h, D})$ computed in this manner --- we consider a 1D spin system of size $N = 40$ and circuits designed to prepare the ground state of a commuting local Hamiltonian (see figure caption for details). The plots in Fig.~\ref{fig:spinsentropic} show the bounds for MPO ansatzes with different bond dimensions $D$ plotted against the circuit depth $d$ for circuits with noise rates $p = 3\%$ [Fig.~\ref{fig:spinsentropic}(a)], $p = 5\%$ [Fig.~\ref{fig:spinsentropic}(b)], $p = 10\%$ [Fig.~\ref{fig:spinsentropic}(c)], and $p = 20\%$ [Fig.~\ref{fig:spinsentropic}(d)].  However, for the lowest noise rate $p = 3\%$, the circuit dual bounds at intermediate depths are trivial i.e. lower than the ground state energy of $H$ --- this can be attributed to the fact that the intermediate depth regime is the regime where the MPO ansatz is least representative. For shorter depths, the bond dimension of the Heisenberg picture operator would have not grown very much while for very long depths, the action of the depolarizing noise reduces the bond dimension of the Heisenberg picture operator.

Figure~\ref{fig:spinsentropic} also compares the trace purity-based dual bound to the information content-based bound. However, since the duality-based bound is exactly computable on a classical computer, to make a fair comparison we need to use a certifiable method for computing the information-content based bound. In particular, using Lagrangian duality, the information content-based bound can be reframed in terms of the Gibbs free energy of the problem Hamiltonian i.e.
\begin{align}\label{eq:free_energy_no_cutoff}
\ell^I &= \underset{\rho: I(\rho) \leq N(1-p)^d}{\text{minimize}} \textnormal{Tr}(H\rho)\nonumber\\
&= \underset{\lambda \geq 0}{\text{maximize } } \lambda S_d +G(H, \lambda),
\end{align}
where $S_d = N - N(1-p)^d$ and $G(H, \lambda) = - \lambda \log \textnormal{Tr} \exp(-H/\lambda)$ is the Gibbs free energy of $H$ at temperature $\lambda$. However, since $H$ is generally a many-body Hamiltonian, an accurate evaluation of $G(H, \lambda)$ can only be guaranteed at sufficiently high temperatures \cite{kuwaharaClusteringConditionalMutual2020a}. Thus, instead of evaluating the bound $\ell^I$ in Eq.~\ref{eq:free_energy_no_cutoff}, we introduce a lower bound $\lambda_c$ on the temperature $\lambda$ and evaluate
\begin{align} \label{eq:lbound_temp_ic}
    \ell^I_{\lambda_c} = \underset{\lambda \geq \lambda_c}{\text{maximize } } \lambda S_d +G(H, \lambda),
\end{align}
For spatially local Hamiltonians, $\lambda_c$ can be chosen depending on the norms of the local terms in the Hamiltonian, the dimensionality of the lattice, and the interaction range. In our calculations, we make the choice of $\lambda_c = 8e^3$ --- this is based on Ref.~\cite{kuwaharaClusteringConditionalMutual2020a} which, to the best of our knowledge, provides the only rigorous algorithm that works for evaluating $G(H, \lambda)$ at temperatures above $\lambda_c$. 

We see from Fig.~\ref{fig:spinsentropic} that the dual provides a tighter lower bound on the output than the bound based on just the information content of the output state. The information content-based bounds shown in Fig.~\ref{fig:spinsentropic} are also trivial (i.e.~lower than the ground-state energy) for intermediate \emph{and} short depths --- this is due to the temperature lower bound that needs to be introduced to ensure computability of the Gibbs free energy. We also observe that the separation between the information content-based and circuit dual bounds increases with the bond dimension $D$ as the MPO ansatz becomes more expressive with increasing bond dimension. In the limit of large circuit depth at non-zero depolarizing noise rates, the state of the circuit approaches the completely mixed state, and we observe that both bounds also approach the energy corresponding to the completely mixed state.

\begin{figure}[t]
    \centering
    \includegraphics[width=0.4\textwidth]{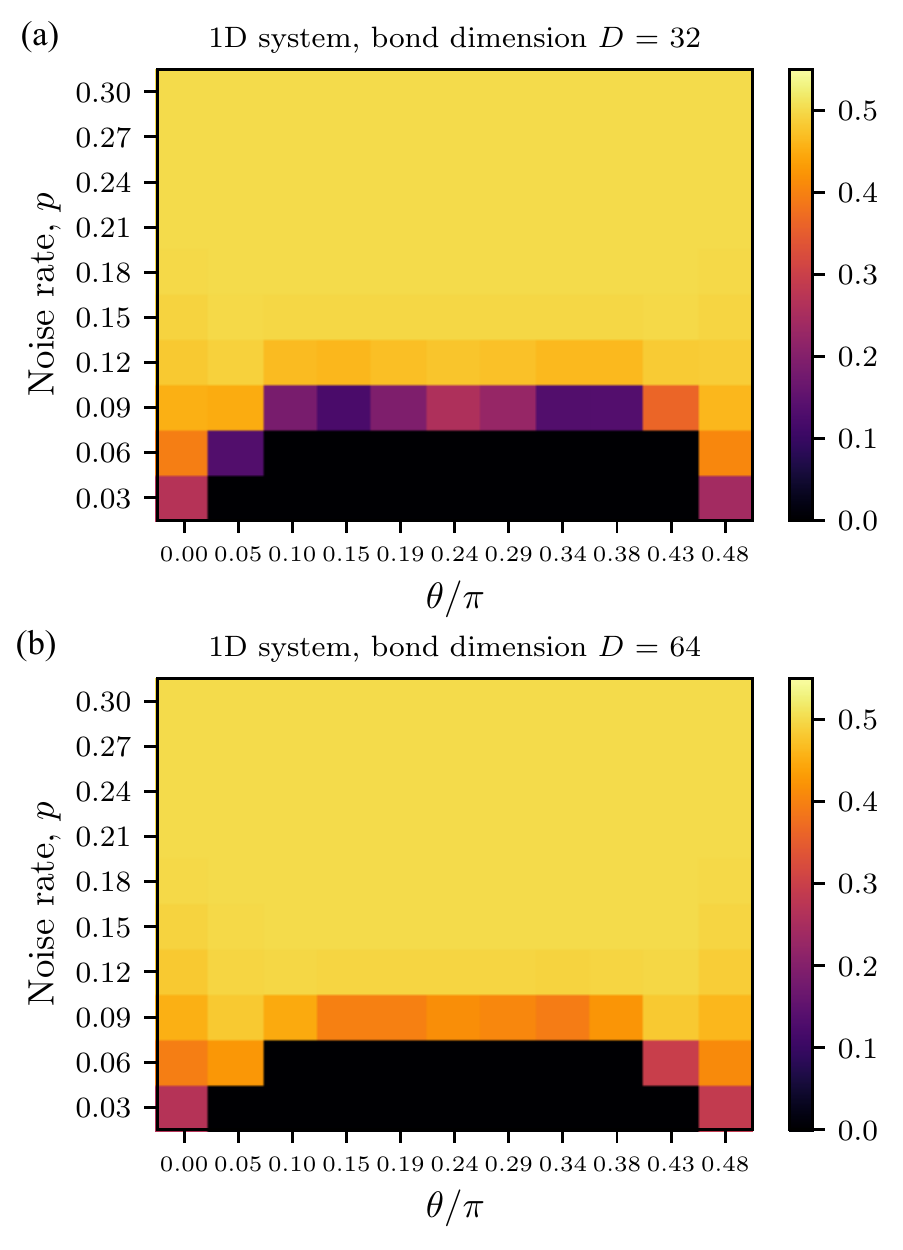}
    \caption{ Trace purity-based dual bounds ($h(\vec{\sigma}^{h, D})$ in Eq.~\ref{eq:duality_bound_trace_purity}) for a 1D system of $N = 32$ spins, the same target Hamiltonian as considered for the results in Fig.~\ref{fig:spinsheis} (see description of Hamiltonian in caption of Fig.~\ref{fig:spinsheis}), and brick-wall quantum circuits [Fig.~\ref{fig:spinsheis}(a)] where two-qubit unitaries are chosen to be $\exp(-i \theta X \otimes X)$ and single-qubit unitaries are independently Haar random. Plots show dual bounds as a function of noise rate $p$ and circuit parameter $\theta$ for circuit depth $d = 25$ and bond dimensions (a) $D = 32$ and (b) $D = 64$. The Hamiltonian is shifted and scaled such that its spectrum is in $[0,1]$.}
    
    \label{fig:spinsparam1D}
\end{figure}

\begin{figure*}[t]
    \centering
    \includegraphics[width=0.8\textwidth]{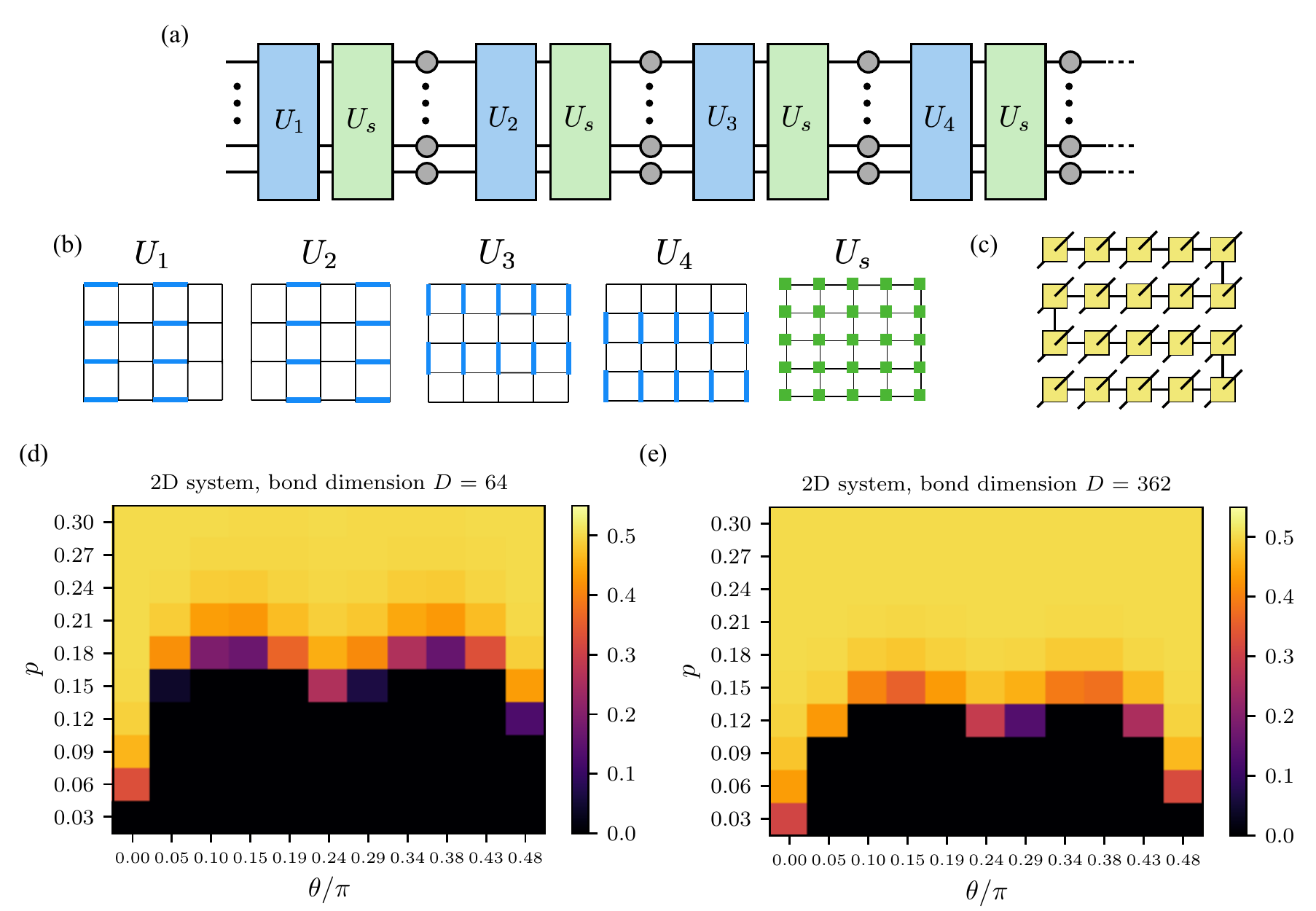}
    \caption{
    (a) Schematic of quantum circuits considered for 2D spin systems on a square lattice: colored boxes indicate unitaries and grey circles depolarizing noise. Each unitary layer consists of two-qubit unitaries $\exp(-i \theta X \otimes X)$ (blue boxes) followed by independently Haar random single-qubit unitaries (green boxes). The first $d/2$ layers serve to increase the entanglement in the state. The remaining layers invert the action of the previous $d/2$ such that, in the absence of noise, the output of the circuit is the ground state of $H =  -\sum_{\langle i, j \rangle} Z_i Z_j $ where $Z_i$ is the Pauli-$Z$ operator for the $i^{\text{th}}$ spin and $\langle i, j \rangle$ indicates nearest-neighbors. (b) Circuits considered have a brick-wall structure: two-qubit unitary layers cycle between gates on odd horizontal edges ($U_1$), even horizontal edges ($U_2$), odd vertical edges ($U_3$), and even vertical edges ($U_4$). Single-qubit gates ($U_s$) are applied on every qubit after every two-qubit gate layer. (c) Structure of the MPO considered for 2D spin systems: yellow squares indicate tensors at each site in the 2D lattice and lines emerging from them indicate tensor indices. Horizontal lines indicate bond indices with dimension $D$ and diagonal lines indicate physical indices. (d,e) Trace purity-based dual bounds ($h(\vec{\sigma}^{h, D})$ in Eq.~\ref{eq:duality_bound_trace_purity}) for the ground state energy of the target Hamiltonian $H$ as a function of noise rate $p$ and circuit parameter $\theta$ for a system of $N = 36$ spins arranged in a $6 \times 6$ lattice, circuit depth $d = 32$ and MPO bond dimensions (d) $D = 64$ and (e) $D = 362$. The Hamiltonian is shifted and scaled such that its spectrum is in $[0,1]$.}
    \label{fig:spinsparam2D}
\end{figure*}

Finally, we demonstrate that the dual bounds are able to capture the extent of entanglement being generated in a circuit. In the brick-wall quantum circuits we consider, the two-qubit gates in the circuit $\exp(-i \theta X \otimes X)$ are parametrized by an angle $\theta$ which controls the entanglement being produced --- for example, at $\theta = 0, \pi/2$, there is no entanglement at all. Fig.~\ref{fig:spinsparam1D} shows the bounds as a function of the angular parameter $\theta$ and the noise rate $p$, for constant bond dimensions $D$ and circuit depths $d$, for a 1D system of $N = 32$ spins and Fig.~\ref{fig:spinsparam2D}(d,e) show the same for a 2D system of $N = 36$ spins in a $6 \times 6$ lattice. For the 2D system, we consider the MPO ansatz to have a `snake-like' bond structure on the 2D lattice [Fig.~\ref{fig:spinsparam2D}(c)] --- such a snake-like structure is a numerically convenient approach for  performing TEBD for 2D systems. This ansatz is useful for moderate system sizes but due to gates along the vertical edges of the lattice the bond dimension required grows rapidly. For larger system sizes, we expect that a tensor network ansatz that matches the architecture of the circuit \cite{panContractingArbitraryTensor2020, markovSimulatingQuantumComputation2008, lubaschAlgorithmsFiniteProjected2014} would give better bounds. For both the 1D and 2D systems, the target Hamiltonians are shifted and scaled such that the ground state energies are zero and any bounds lower than zero are considered trivial and represented as zero in the plots --- the black regions in the plots thus correspond to trivial bounds. We observe that, near $\theta = 0, \pi/2$, where the entanglement is small, the MPO ansatz of constant bond dimension used for the bounds is able to capture it and we obtain non-trivial bounds for small noise rates $p \approx 6\%$. For values of $\theta$ away from these limits, the region of triviality is larger but non-trivial dual bounds can still be obtained for higher noise rates.

\subsection{Non-depolarizing noise models}
Up until now, we have modeled the noise present in the circuit as depolarizing. However, noise in several experimental systems might have a more complex structure. In this subsection, we consider these other noise models and show that the duality-based bounding procedure can be reformulated slightly to provide informative bounds even without the assumption of the depolarizing noise. The only assumption that we make that the noise channel under consideration has a full Kraus rank (i.e.~the Kraus operators describing the noise channel span the entire space of singe qubit operators). This assumption could be seen as a reasonable physical assumption for sufficiently generic noise models --- if the Kraus operators are interpreted effectively as operators that randomly act on the qubit when it is experiencing noise, the Kraus operators not spanning the full space of linear operators would mean that the noise is special and does \emph{not} apply an entire subspace of operators on the qubit. Nevertheless, for channels that do not have full Kraus rank, the methods presented in this section do not apply and we leave it as an open problem for future work.

First, consider noise channels that are unital and primitive (i.e.~have identity as a fixed point) --- in this case, the noise channel $\mathcal{N}$ with noise rate $p$ can always be expressed as
\begin{align}\label{eq:unital_primitive}
\mathcal{N}(\rho) = (1 - p) \rho + V\bigg(\sum_{P \in \{X, Y, Z\}}p_P P U \rho U^\dagger P \bigg)V^\dagger,
\end{align}
for a single-qubit unitaries $U, V$ and $p_X, p_Y, p_Z \in (0,1)$ with $p = p_X + p_Y + p_Z < 1$. In this case, as detailed in appendix, it follows from an application of a corollary of Ref.~\cite{hirche2022contraction} that Lemma~\ref{lemma:depolarizing_entropy_bounds} can be extended to this class of channels with the noise rate being chosen as $\min(p_X, p_Y, p_Z)$.
\begin{lemma}[Follows from Corollary 5.6 of Ref.~\cite{hirche2022contraction}]
Suppose $\rho_t$ is the quantum state on $N$ qubits obtained from an initial pure state after applying $t$ unitaries followed by single qubit noise channels of the form of Eq.~\ref{eq:unital_primitive}, then
\begin{align}
P_\textnormal{tr}(\rho_t) := \textnormal{Tr}(\rho_t^2) \leq 2^{-N(1 - (1 - \min(p_x, p_y, p_z))^t)}.
\end{align}
\end{lemma}

The case of non-unital noise channels is more complex, and it is not possible to get architecture independent bounds on the entropy or trace-purity of the time-dependent state of the quantum circuit. We restrict ourselves to the case where the non-unital noise channel under consideration has a unique fixed point $\tau$ --- this noise channel, then, tends to drive the output of a quantum circuit on $N$ qubits to the state $\tau^{\otimes N}$. Therefore, instead of using a trace-purity constraint while formulating the dual, as we have for unital noise channels, we instead use a constraint on $\norm{\rho_t - \tau^{\otimes N}}_F^2$, where $\norm{X}_F^2 = \text{Tr}(X^\dagger X)$ is the Frobenius norm of $X$. To obtain a simple analytical upper bound on $\norm{\rho_t - \tau^{\otimes N}}_F^2$, we assume that the two-qubit gates in the quantum circuit are diagonal in the basis of eigenvectors of $\tau$. This is the case, for e.g., if the non-unital noise channel is amplitude damping and the two-qubit gates used in the quantum circuit are all control phase gates. We point out that if prior information about the fixed point of the noise channel is known (for e.g. from a previous noise tomography), then a universal gate set can always be chosen such that the two-qubit gates satisfy this requirement. 

More concretely, suppose that the unitary in time step $t$ has single-qubit gates $V_{t, \alpha}^{(1)}$. We also assume that the noise channel, $\mathcal{N}$, is non-unital and has a full Kraus rank, which we expect to be true for noisy systems if the noise is sufficiently generic. Denoting the fixed point of $\mathcal{N}$ by $\tau$, we show in appendix that if $\mathcal{N}$ has a full Kraus rank, then 
\begin{align}\label{eq:noise_channel}
\mathcal{N} = \Lambda_{\tau, q} {\mathcal{N}'},
\end{align}
for some channel $\mathcal{N}'$ which also has $\tau$ as a fixed point and $\Lambda_{\tau, q}(X) = (1 - q) X + q\text{Tr}(X) \tau$ for $0 < q < 1$. The parameter $q$ can be interpreted as the probability with which the noise channel traces out the qubit and replaces it with $\tau$, and it can be computed by solving the following semi-definite program
\begin{equation}\label{eq:opt_problem}
\begin{aligned}
\max_{q} \quad & q\\
\textnormal{s.t.} \quad & \Phi_\mathcal{N} - q I\otimes \tau \succeq 0,\\
  &q > 0, q < 1.    \\
\end{aligned},
\end{equation}
where $\Phi_{\mathcal{N}}$ is the Choi state corresponding to $\mathcal{N}$. For such noise channels and unitary circuits,
\begin{lemma}\label{lemma:distance_general_noise}[Follows from Lemma 1 of Ref.~\cite{stilckfrancaLimitationsOptimizationAlgorithms2021}]
    Suppose $\rho_t$ is the quantum state on $N$ qubits obtained from an initial state $\rho_0$ after applying $t$ unitaries followed by single qubit noise channel $\mathcal{N}$ of the form of Eq.~\ref{eq:noise_channel}, then it follows that
    \begin{align}\label{eq:final_quantum_relative_entropy}
    &D(\rho_t |\hspace{-0.05cm}|\tau^{\otimes N}) \leq D(\rho_0 |\hspace{-0.05cm}|\tau^{\otimes N}) (1 - q)^t + \nonumber\\
    &\qquad \quad 2\sum_{s = 0}^{t - 1} \sum_{\alpha}(1 - q)^{t - s} \log_2(\smallnorm{\tau^{-1/2}V_{\alpha, s}\tau V_{\alpha, s}^\dagger \tau^{-1/2}})
    \end{align}
    where $D(\rho_1 |\hspace{-0.005in}|\rho_2 ) = \textnormal{Tr}[\rho_1 \log_2 \rho_1 - \rho_1 \log_2 \rho_2]$ is the quantum relative entropy between $\rho_1$ and $\rho_2$.
\end{lemma}
\noindent We outline a full proof of this lemma in appendix B. To translate the upper bound on the quantum relative entropy to an upper bound on  $\norm{\rho_t - \tau^{\otimes N}}_F$, we note that
\begin{align*}
    \norm{\rho_t - \tau^{\otimes N}}_F^2 \leq 2 \norm{\rho_t - \tau^{\otimes N}}_1 \leq \big[2D(\rho_t |\hspace{-0.05cm}|\tau^{\otimes N})\big]^{1/2},
\end{align*}
where we have used that $\norm{O}_2^2 \leq \norm{O}_1 \norm{O}$ and the Pinsker's inequality $\norm{\rho_1 - \rho_2}_1 \leq \sqrt{\frac{1}{2}D(\rho_1 |\hspace{-0.05cm}|\rho_2)}$. We remark that our bounds become trivial (i.e.$\to \infty$) when $\tau$ is not full rank which would also imply, by the quantum Perron-Frobenius theorem, that the noise channel doesn't have a full Kraus rank.

Following the same procedure as in Section \ref{duality-based-bounds}, we can now formulate the following optimization problem for the energy at the output quantum circuit.
\begin{align} \label{eq:n_qubit_primal_problem_general_noise}
    \underset{{\rho_1, \rho_2 \dots \rho_d} \in \mathcal{S}}{\text{minimize}} \quad & \text{Tr}(H\rho_d) \nonumber\\
    \textrm{subject to} \quad & \rho_t = \mathcal{E}_t(\rho_{t-1}), \ t \in \{1, \ldots, d \},\nonumber \\
    & \norm{\rho_t - \tau^{\otimes N}}_F^2\leq d_t^2, \ t \in \{1, \ldots, d \},\numberthis
\end{align}
where, instead of a trace purity bound as in Eq.~\ref{eq:n_qubit_primal_problem}, we use an upper bound on the Frobenius norm distance between $\rho_t$ and $\tau^{\otimes N}$ with $d_t$ given by lemma \ref{lemma:distance_general_noise}. Again, introducing the dual variables $\sigma_1, \sigma_2 \dots \sigma_d \in \mathcal{M}((\mathbb{C}^2)^{\otimes N})$ and $\lambda_1, \lambda_2 \dots \lambda_t \geq 0$, we can construct the Lagrangian

\begin{align}
\label{eq:lag_noise}
&\mathcal{L}(\vec{\rho}, \vec{\sigma}, \vec{\lambda}) = \text{Tr}[H\rho_d] + \sum_{t = 1}^d \text{Tr}\bigg[\sigma_t\big(\rho_t - \mathcal{E}_t(\rho_{t - 1})\big) \bigg] + \nonumber \\
&\qquad\qquad \sum_{t = 1}^d \lambda_t\bigg[\norm{\rho_t - \tau^{\otimes N}}_F^2 - P_t \bigg].
\end{align} 
Minimizing the Lagrangian over $\vec{\rho}$ and then maximizing it over $\vec{\lambda}$ yields the dual function $h(\vec{\sigma})$ 
\[
h(\vec{\sigma}) = -\textnormal{Tr}(\sigma_1 \mathcal{E}_1(\rho_0)) + \sum_{t = 1}^d\bigg[\textnormal{Tr}(H_t \tau^{\otimes N}) - d_t\sqrt{\text{Tr}(H_t^2)}\bigg],
\]
where, again, $H_d = H + \sigma_d$ and $H_t = \sigma_t - \mathcal{E}_{t + 1}^\dagger(\sigma_{t + 1})$ for $t \in \{1, 2 \dots d - 1\}$. Similar to the case in the previous sub-sections, due to Lagrangian duality, $g(\vec{\sigma})$ evaluated at any $\vec{\sigma}$ provides a lower bound on the energy at the output of the circuit. Following the strategy in the previous sections, we again evaluate $g(\vec{\sigma})$ at $\vec{\sigma}^h$ given by a TEBD algorithm in the Heisenberg picture (Eq.~\ref{eq:heis_tebd}). As an example, the bounds obtained in a 1D circuit on $N = 30$ qubits, at a noise rate of $3\%$, with noise modelled by the non-unital replacement channel $\mathcal{N}(X) = (1 - q) \rho + q \textnormal{Tr}(X) \tau$, where we assume $\tau = I/2 + \epsilon Z$. Here $\epsilon$ controls how ``non-unital" the noise channel is --- $\epsilon = 0$ corresponds to the previously studied case of a depolarizing noise channel and in the limit of $\epsilon \to \pm 1/2$, we obtain an amplitude damping channel (which does not have a full Kraus rank). We also compute the bounds obtained by considering only the quantum relative entropy of the circuit output ($\rho_d$) with respect to the noise channel fixed point $\tau^{\otimes N}$ as given by Eq.~\ref{eq:final_quantum_relative_entropy} --- to obtain this, we note that given an upper bound on $D(\rho_d |\hspace{-0.05cm}|\tau^{\otimes N})$, we can translate it to an upper bound on $\norm{\rho_d - \tau^{\otimes N}}_1$ via the Pinkser's inequality i.e.
\[
\norm{\rho_d - \tau^{\otimes N}}_1 \leq \sqrt{\frac{1}{2} D(\rho_d |\hspace{-0.05cm}|\tau^{\otimes N})},
\]
and therefore the expected energy $\textnormal{Tr}(H \rho_d)$ at the circuit observable can deviate from $\tau^{\otimes N}$ by at-most $\norm{H}\smallnorm{\rho_d - \tau^{\otimes N}}_1$. Thus we obtain the lower bound 
\begin{align}
&\textnormal{Tr}(H\rho_d) \geq \textnormal{Tr}(H\tau^{\otimes N}) - \norm{H}\smallnorm{\rho_d - \tau^{\otimes N}}_1, \nonumber \\
&\qquad \ \ \ \ \ \geq \textnormal{Tr}(H\tau^{\otimes N}) -\norm{H}\sqrt{\frac{1}{2}D(\rho_d |\hspace{-0.05cm}|\tau^{\otimes N})}.
\end{align}
We find that using the duality-based bound continues to give informative bounds which are significantly better compared to the bounds attained by just accounting for the distance between the noisy output state and the noise-channel fixed point $\tau^{\otimes N}$. Specifically, as the noise channel becomes increasingly non-unital, then the dual formulation continues to provide non-trivial bounds since it accounts for the circuit architecture, and the bounds attained without accounting for the circuit architecture become trivial for even slightly non-unital channels.

\begin{figure}
    \centering
    \includegraphics[scale=0.5]{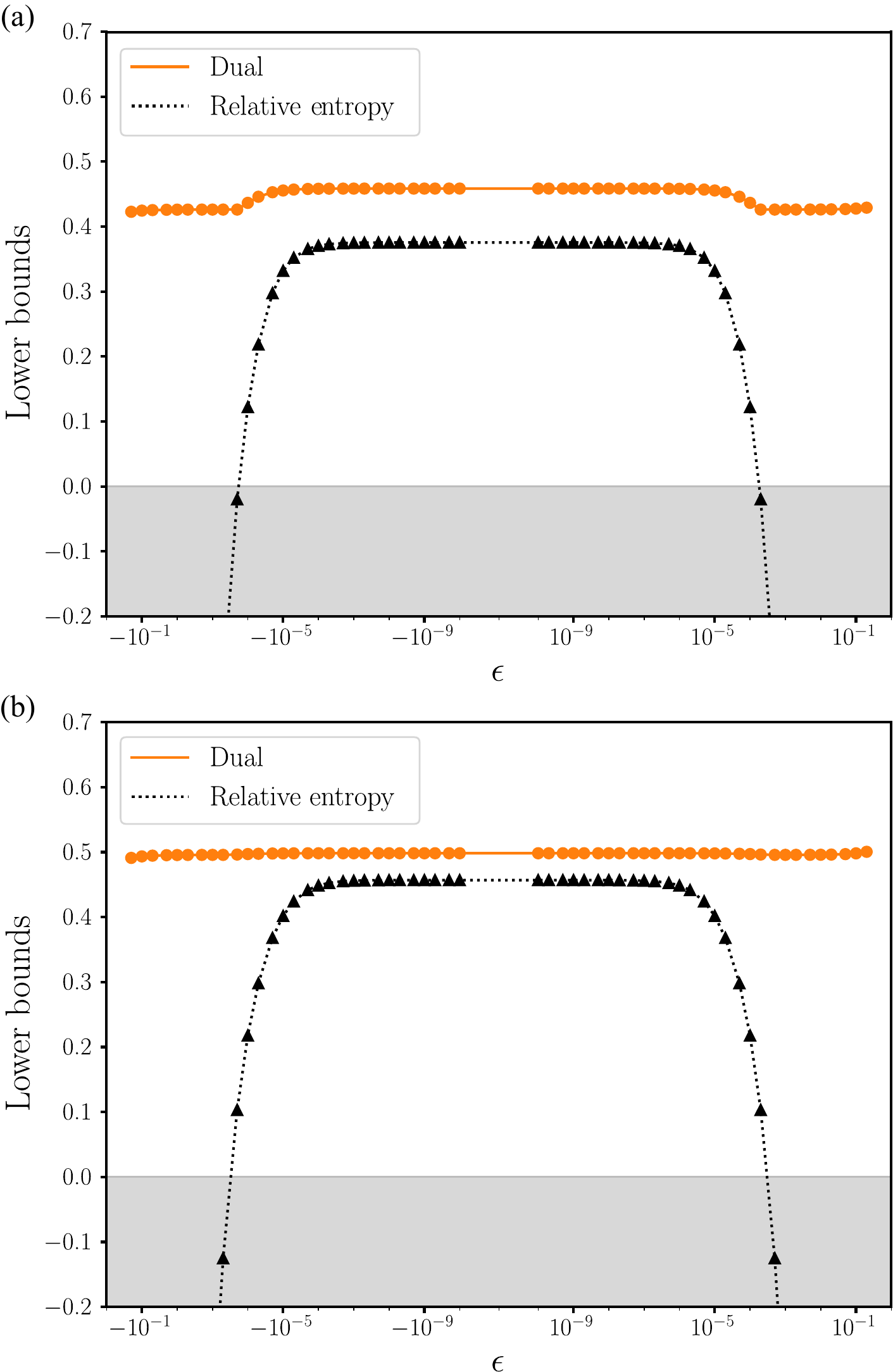}
    \caption{Bounds on the output energy for the case of a non-unital noise channel. The structure of the circuits are identical to those considered in Fig.~\ref{fig:spinsheis}(a) with parameters $N = 32$, $d = 102$ and $\theta = 0.1$. The local noise after each layer of unitaries is given by $\mathcal{N}(X) = (1 - q) X + q\textnormal{Tr}(X) \tau_\epsilon$, with $\tau_\epsilon = (1/2 + \epsilon) \ket{0}\!\bra{0} + (1/2 - \epsilon) \ket{1}\!\bra{1}$, with $q = 0.03$ in (a) and $q = 0.05$ in (b). In orange, we show the bounds derived on the output energy from our dual problem with an MPO ansatz with $D = 128$ for the dual variables $\sigma_1, \sigma_2 \dots \sigma_d$. The black lines show the lower bound $\textnormal{Tr}(H \tau^{\otimes N}) - ||H|| ||\rho_d - \tau^{\otimes N}||_1$ which is obtained by disregarding the circuit architecture.}
    \label{fig:non-depolarizing-noise}
\end{figure}

\section{Using duality with Information content} \label{sec:information}

In the previous sections, we have investigated the impact of noise using the trace purity as a measure of the mixedness in the noisy circuit. The trace purity-based dual function Eq.~\ref{eq:duality_bound_trace_purity} contains terms with Frobenius norms $\sqrt{\text{Tr}(H_t^2)}$ which, in the worst case, could grow exponentially with the system size $N$. Hence, the trace purity-based dual tends to become trivial in the limit of large system size and intermediate circuit depths. An alternative better conditioned purity measure is the information content based on the Von-Neumann entropy, $I(\rho) = N - S(\rho) = N + \text{Tr}(\rho \log_2 (\rho))$. In this section, we formulate a duality-based bound using the information content as a purity measure. However, as we illustrate below, the duality bound here is harder to compute than the one based on trace-purity for general spin model --- to still gauge the efficacy of this bound, we numerically study in the simpler but physically relevant case of Gaussian fermions. While our results are suggestive that using the information-content based bounds could be useful for spin models, the associated dual function is harder to compute classically --- we leave it as an open problem to develop classical algorithms to compute informative bounds using this strategy for spin models.

Considering the information content, the free energy defined in Eq.~\ref{eq:generalized_free_energy} becomes the Gibbs free energy with an offset, 
\begin{align} \label{eq:ic_free_energy}
    \mathcal{F}_{\mathcal{S}, P}(H, \lambda) &= \inf_{\rho \succeq 0, \text{Tr}(\rho) = 1}\bigg( \textnormal{Tr}[H\rho] + \lambda I(\rho)\bigg), \nonumber\\
    &=N\lambda + \inf_{\rho\succeq 0, \textnormal{Tr}(\rho) = 1} \big(\text{Tr}(H\rho) - \lambda S(\rho) \big), \nonumber\\
    &= N \lambda -\lambda \log\text{Tr}\exp\left(-H/\lambda\right),
\end{align}
which together with Eq.~\ref{eq:dual_function} yields,
\begin{align*}\label{eq:ic_dual_function}
\tilde{g}(\vec{\sigma}, \vec{\lambda}) &= -\textnormal{Tr}[\rho_0 \mathcal{E}^\dagger_1(\sigma_1)] \\ 
&+ \sum_{t = 1}^d \bigg(-\lambda_t \log\text{Tr}\exp\left(-H_t/\lambda_t\right) + \lambda_t (N - I_t)\bigg), \numberthis
\end{align*}
where $I_t = N (1-p)^t$ is the analytical bound on the information content under depolarizing noise defined in Lemma 1. 

To benchmark the performance of the information content-based dual, we consider Gaussian fermionic systems where the dual function Eq.~\ref{eq:ic_dual_function} can be computed exactly. We study $N$ fermions arranged on a lattice and choose $H$ to be a quadratic Hamiltonian,
\[
H = i\sum_{\substack{\alpha, \alpha', x, x'}} h_{x, x'}^{\alpha, \alpha'} c_x^\alpha c_{x'}^{\alpha'},
\]
where $c_x^{1}, c_x^{2}$ are the Majorana operators for the fermion at point $x$ on the lattice, and $h_{x, x'}^{\alpha, \alpha'}$ are real numbers specifying $H$. We additionally assume the unitaries in the circuit that prepares the ground state of $H$ from an initial vacuum state to be Gaussian unitaries.

\begin{figure}[t]
    \centering
    \includegraphics[scale=1.0]{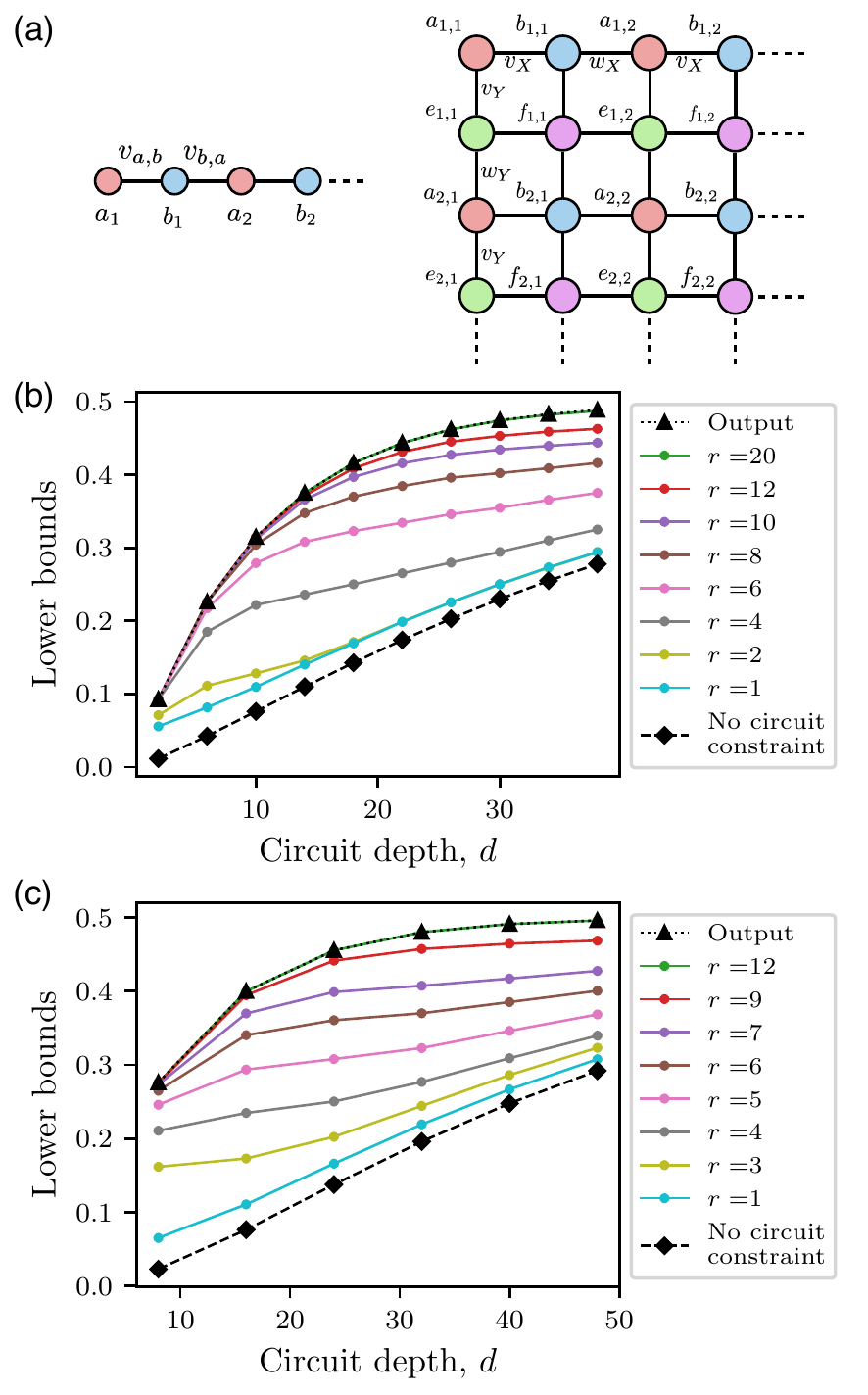}
    \caption{(a) Schematic of SSH model quadratic fermionic Hamiltonians with alternating hopping strengths in 1D and 2D. (b,c) Comparison of information content-based dual bounds with and without circuit constraints, and the output energies of noisy Gaussian circuits for systems consisting of (b) $N = 48$ fermions arranged in a 1D lattice, (c) $N = 49$ fermions arranged in a $7  \times 7$ 2D lattice. Dual bounds are shown for ansatzes with varying interaction range $r$. The horizontal axis represents the depth $d$ of a Gaussian brick-wall circuit that outputs the ground state of the SSH model. Fermions are independently subject to depolarizing noise with probability $p = 5\%$ after every unitary layer. The Hamiltonians are shifted and scaled such that their spectrum is in $[0,1]$.}
    \label{fig:fermions}
\end{figure}

Since both Gaussian unitaries and the depolarizing channel map a quadratic Hermitian operator to another quadratic Hermitian operator, Proposition 1 indicates that the dual function is maximized for $\sigma_t$ which themselves are quadratic Hermitian operators. This motivates the following ansatz for $\sigma_t$
\[
\sigma_t = i\sum_{\substack{\alpha, \alpha', x, x' \\ d(x, x') \leq r}} s_{x, x'; t}^{\alpha, \alpha'} c_x^\alpha c_{x'}^{\alpha'},
\]
for real $s_{x, x'; t}^{\alpha, \alpha'}$. In our study, we restrict $\sigma_t$ to be local operators with interaction range $r$ while maximizing $g(\vec{\sigma}, \vec{\lambda})$ to obtain the lower bound --- when $r\sim$ lattice size, we expect to obtain the best possible lower bound but since the ansatz always includes the point $\vec{\sigma} = 0$, we expect from Proposition 2 to obtain a bound better than that predicted by only considering the information content of the output state, even for small $r$. Choosing $\sigma_t$ to be quadratic Hermitian operators allows for exact, classically efficient computation of the Gibbs free energy terms in the dual function and, furthermore, even the circuit output can be computed exactly by considering the covariance matrix describing the state --- see appendix B for details. We obtain bounds by maximizing $g(\vec{\sigma}, \vec{\lambda})$ through a gradient-based local optimization algorithm (L-BFGS-B), starting from the initial point where $\sigma_t$ are chosen to be the Heisenberg picture evolution of $-H$, but projected on to the space of quadratic fermionic Hamiltonians with interaction range $r$ after each time step --- much like the compression into MPOs of bond dimension $D$ in Eq.~\ref{eq:heis_tebd}.

Figure \ref{fig:fermions} shows a numerical study of the bounds that we obtain --- we consider systems with $\sim 50$ fermions arranged both on 1D [Fig.~\ref{fig:fermions}(b)] and 2D lattices [Fig.~\ref{fig:fermions}(c)] and experiencing depolarizing noise at a rate of 5$\%$. $H$ is chosen to be a SSH model, nearest neighbor Hamiltonian with alternating hopping strengths [Fig.~\ref{fig:fermions}(a)]. For the 1D benchmarks, we choose
\begin{subequations}
    \begin{align}
H = &\sum_{x} \bigg(v_{a, b}  a_x^{\dagger}b_x + v_{b, a} b_x^\dagger a_{x + 1} + \text{h.c.}\bigg),
    \end{align}
and for 2D benchmarks, we choose
    \begin{align*}
     H = &\sum_{\substack{x, y}} \Bigg( \sum_{\substack{p, q \in \\ \{(a, b), (e, f)\} }}\bigg(v_{p, q}  p_{x,y}^{\dagger}q_{x, y} + v_{q, p} q_{x, y}^\dagger p_{x + 1, y}  + \text{h.c.}\Bigg)  \nonumber \\
     & +\sum_{\substack{p, q \in \\ \{(a, e), (b, f)\} }}\bigg(v_{p, q}  p_{x,y}^{\dagger}q_{x, y} + v_{q, p} q_{x, y}^\dagger p_{x, y + 1} + \text{h.c.}\Bigg) \Bigg), \numberthis
    \end{align*}
\end{subequations}
where we choose $v_{a, b} = v_{e f} = v_x, v_{b, a} = v_{f, e} = w_x, v_{a, e} = v_{b, f} = v_y$ and $v_{e, a} = v_{f, b} = w_y$. For the numerical studies shown in Fig.~\ref{fig:fermions}, $H$ is shifted and scaled such that the ground and highest excited state energies are zero and one, respectively. We consider circuits of depth $d$ consisting of two-mode Gaussian unitaries arranged in a brick-wall layout, where the first $d/2$ layers are composed of randomly generated two-mode Gaussian unitaries that serve to increase the entanglement in the state. The remaining layers invert the action of the previous $d/2$ such that, in the absence of noise, the output state is the initial state, which is chosen to be the ground state of $H$. In Fig.~\ref{fig:fermions}, for comparison, we also include the exact output of the noisy Gaussian circuit, as well as the bound obtained by neglecting the circuit constraints and only considering the information content of the output state. As expected, we find that on accounting for the circuit constraint, we obtain bounds that are more representative of the output. We also observe that the dual bounds get closer to the output as the dual ansatz's interaction range $r$ increases, since the ansatz becomes more expressive. 

\section{Conclusion and outlook}

In conclusion, we demonstrate a method to rigorously lower bound the performance of any given quantum circuit subject to a constant rate of depolarizing noise. We achieve this by constructing a Lagrangian dual specific to the circuit, which takes into consideration not only the decreasing purity of the state through the circuit due to noise, but also the details of the gates in the circuit, allowing the study of the effect of entanglement generation in the circuit that can worsen the detrimental effects of noise. We presented numerical studies in spin systems and showed that it is possible to efficiently calculate circuit-specific lower bounds that are tighter than bounds obtained by just considering the information content of the output state. We provided an interpretation of the trace purity-based dual evaluated at dual variables obtained from TEBD in the Heisenberg picture in terms of the compression errors. We also showed that the dual can be formulated in terms of the information content of the state instead of trace purity --- we computed information content-based circuit dual bounds for Gaussian fermionic systems where the Gibbs free energy can be computed exactly. 

Our method opens the door to promising avenues of future research. Larger-scale tensor-network numerics can allow us to also study higher dimensional circuits, with a number of qubits reaching state-of-the-art experiments. Numerical optimization algorithms can be explored for obtaining better lower bounds than those from evaluating the dual function at a specific point. Moreover, extensions of the methods of this paper to continuous-time would better capture the experimental system, and even allow us to apply this method for understanding quantum adiabatic algorithms \cite{albashAdiabaticQuantumComputation2018}. Finally, the Lagrangian dual formulation of lower bounds, apart from being a numerical tool, could also shed rigorous theoretical insights in understanding resilience of quantum circuit architectures to noise.

\section*{Acknowledgements}
We thank J. Ignacio Cirac, Mari Carmen Ba$\tilde{\text{n}}$uls and Guillermo Gonz\'{a}lez-Garc\'{i}a for helpful discussions. R.T. acknowledges a Max Planck Harvard Research Center for Quantum Optics(MPHQ) Postdoctoral Fellowship. 

\onecolumngrid
\newpage

\appendix

\section{Proof of proposition 4}\label{app:proof_prop_4}

We first recall the main result of Ref.~\cite{gonzalez-garciaErrorPropagationNISQ2022} which analyzed error propagation in a family of random quantum circuits. Specifically, they considered brick-wall quantum circuits of depth $d$ (assumed to be even) with unitaries $U_1, U_2 \dots U_d$, where
\[
U_{d/2 + 1} = U_{d/2}^\dagger, U_{d/2 + 2} = U_{d/2 - 1}^\dagger, U_{d/2 + 3} = U_{d/2 - 2}^\dagger \dots U_d = U_1^\dagger.
\]
The unitaries $U_1, U_2 \dots U_{d/2}$ are chosen randomly depending on the circuit architecture. We will specifically consider the 1D case, where $U_1, U_3, U_5 \dots$ are formed by applying random 2-qubit gates between qubits $(1, 2), (3, 4) (5, 6) \dots$ and $U_2, U_4, U_6 \dots$ are formed by applying random 2-qubit gates between qubits $(2, 3), (4, 5), (6, 7) \dots$. All the two qubit gates are chosen independently at random from an ensemble that forms a 2-design. Furthermore, we consider the noisy setting where depolarizing noise with probability $p$ is applied to each qubit after every unitary layer. Ref.~\cite{gonzalez-garciaErrorPropagationNISQ2022} establishes the following result characterizing the average energy of the output state for a 2-local Hamiltonian. While the result of Ref.~\cite{gonzalez-garciaErrorPropagationNISQ2022} holds for arbitrary two-local Hamiltonians, we will only consider the Hamiltonian $H = -\sum_{i=1}^N Z_i$.
\begin{lemma}[Ref.~\cite{gonzalez-garciaErrorPropagationNISQ2022}]
The expectation value $E$ of the Hamiltonian $H = -\sum_{i=1}^N Z_i$ with respect to the the output state of a circuit chosen randomly from the ensemble described above satisfies, 
\begin{align*}
\textnormal{Prob}\left( \left| E + N(1-p)^{\Omega(d^2)} \right| \leq \alpha_0 \sqrt{N} \right) \geq 1 - 2e^{-2\alpha_0^2 / 2d^2}. 
\end{align*}
\end{lemma}

\noindent\textit{Proof of proposition 4:} Consider the Hamiltonian $H = -\sum_{i = 1}^{N} Z_i $, and choose the two qubit gates to be a Haar-random Clifford gate --- since Haar-random Clifford gates form a 2-design \cite{dankertExactApproximateUnitary2009}, we can use lemma A.1. $H$ satisfies $\text{Tr}(H) = 0$ and has operator norm $\Vert H \Vert = N$. We evaluate the dual function at the dual variables obtained by Heisenberg picture evolution of $-H$,
\[
\sigma_d = -H, \sigma_t = -\mathcal{E}_{t + 1}^\dagger \mathcal{E}_{t + 2}^\dagger \dots \mathcal{E}_d^\dagger(H).
\]
Note that for Clifford circuits, $Z_i$ will be mapped to a single Pauli string \cite{aaronsonImprovedSimulationStabilizer2004a}, which is expressible as an MPO with bond dimension 1, and consequently $\sigma_t$ will have a bond-dimension of at-most $N$. Since the Heisenberg picture evolution can be captured exactly with a MPO ansatz of bond dimension $O(N)$, the dual bound $\ell^\text{dual}_D$ with $D \leq O(N)$ is exactly equal to the energy $E$ of the output of the noisy quantum circuit. Now, from lemma A.1, it follows that there must exist at least one 1D circuit such that
\[
\ell^\text{dual}_D = -N(1 - p)^{\Omega(d^2)} + O(\sqrt{N}),
\]
with $D \leq O(N)$.

We now consider the lower bound $\ell^I$ obtained by only considering the information content of the state at the output of the circuit and neglecting the circuit constraints:
\begin{align*}
    \ell^I = \underset{ \substack{\rho:  I(\rho) \leq N(1-p)^d}}{\textnormal{minimize}} \  \textnormal{Tr}(H\rho).
\end{align*}
A negative upper bound on $\ell^I$ which converges to $\text{Tr}(H/2^N) = 0$ exponentially with $d$ can be obtained by computing $\text{Tr}(H\rho)$ at 
\[
\rho = \bigg(p_d\ket{0}\bra{0} + (1-p_d) \frac{I}{2}\bigg)^{\otimes N}\ \text{where } p_d = (1-p)^d.
\]
Note that this $\rho$ satisfies $I(\rho) = N(1-p)^d$, and consequently $\text{Tr}(H\rho)$ is an upper bound on $\ell^I$. For $H = \sum_{i = 1}^N Z_i$, we then obtain $\text{Tr}(H\rho) = -N(1-p)^d$ which implies $\ell^I \leq -N(1-p)^{O(d)}$. \hfill $\square$
\section{Non-depolarizing noise channels}
In this subsection, we obtain bounds on trace purity in the presence of non-depolarizing noise channels. The bounds presented here are already contained in or can be straightforwardly obtained from existing results (for e.g.~in Refs.~\cite{de2023limitations, stilckfrancaLimitationsOptimizationAlgorithms2021, hirche2022contraction}). We include this appendix for a self contained derivation of the results used in the main text.

\emph{Unital noise channels}. We first consider unital noise channels that are also primitive. It is a standard result from the characterization of qubit channels that such a noise channel can be expressed as
\begin{align}\label{eq:app_unital_noise_channel}
\mathcal{N}(X) = (1 - p) X  + \sum_{a \in \{x, y, z\}} p_a \mathcal{U} (A \mathcal{V}(X) A),
\end{align}
where $p = p_x + p_y + p_z$ can be considered to be the noise rate, and $\mathcal{V}(\cdot) = V \cdot V^\dagger$, $\mathcal{U}(\cdot) = U \cdot U^\dagger$ are unitary channels with $V, U$ being unitaries. Furthermore, if $\mathcal{N}$ to be primitive (i.e.~have $I/2$ as the unique fixed point), then $p_x, p_y, p_z > 0$. To obtain a bound on the trace purity of the qubit's state in a quantum circuit impacted by such a noise, we need the following lemma from Ref.~\cite{hirche2022contraction}. Here, $\smallnorm{\cdot}_{2\to 2}$ of a super-operator refers to the Schatten-2 norm i.e.
\[
\smallnorm{\mathcal{E}}_{2\to 2}^2 = \sup_{X \neq 0} \frac{\textnormal{Tr}(\mathcal{E}(X)^\dagger \mathcal{E}(X) )}{\text{Tr}(X^\dagger X)}.
\]
Furthermore, for $\sigma \succ 0$, $D_2(\rho |\hspace{-0.005in}| \sigma)$ is the sandwiched 2-Renyi divergence that is given by
\[
D_2(\rho|\hspace{-0.005in}| \sigma) = \log_2\text{Tr}\big( \sigma^{-1/2}\rho \sigma^{-1/2}\rho\big).
\]
In particular, note that if $\rho$ is a $N-$qubit density matrix and $\sigma = (I/2)^{\otimes N}$, we obtain that
\[
D_2(\rho|\hspace{-0.005in}| \sigma)  = N + \log \text{Tr}(\rho^2).
\]
\begin{lemma}[Corollary 5.6 from Ref.~\cite{hirche2022contraction}]\label{lemma:hirche}
    Suppose $\Lambda_{\sigma, p}$ is the channel described by ${\Lambda}_{\sigma, q}(X) = (1 - q) X + q \textnormal{Tr}(X) \sigma$. For $\sigma \succ 0$, define $\Gamma_\sigma$ to be the superoperator $\Gamma_\sigma(X) = \sigma X \sigma^{-1}$. Suppose $\mathcal{N}$ be a channel with $\sigma$ as its fixed point, then If $\smallnorm{\Gamma^{-1/2}_\sigma \mathcal{N} \Lambda_{\sigma, q}^{-1}\Gamma^{1/2}_\sigma }_{2\to 2} \leq 1$, then for any $N > 0$ and $N-$qubit density matrix $\rho$
    \[
    D_2(\mathcal{N}^{\otimes N}(\rho) |\hspace{-0.005in}| \sigma^{\otimes N}) \leq \alpha D_2(\rho |\hspace{-0.005in}| \sigma^{\otimes N}),
    \]
    where $\alpha = 2^{2(1 - \smallnorm{\sigma^{-1}}^{-1}) \log_2 (1 - p)/\log_2(\smallnorm{\sigma^{-1}})}$.
\end{lemma}
\noindent\emph{Proof of lemma 2}: We pick $\mathcal{N}$ to be the unital noise channel in Eq.~\ref{eq:app_unital_noise_channel}, and $\sigma = I/2$. With this choice, we have that $\Lambda_{I/2, q}(X) = (1 - q)X + q\text{Tr}(X) I/2 $ and $\Gamma_{\sigma}(X) = 4 X$. Note that $\smallnorm{\Gamma_{I/2}^{-1/2} \mathcal{N}\Lambda_{\sigma, p}^{-1}\Gamma_{I/2}^{1/2}}_{2\to 2} = \smallnorm{ \mathcal{N}\Lambda_{I/2, p}^{-1}}_{2\to 2}$ and that $\Lambda_{I/2, p}^{-1}(X) = (1 - p)^{-1}\big(X - p \text{Tr}(X) I/2 \big)$. Further analysis is simplified in the Pauli basis --- since both $\mathcal{N}$ and ${\Lambda_{I/2, q}^{-1}}$ are unital superoperators, it follows that, if written as $4\times 4$ matrices in the Pauli basis, they have the form
\[
\Lambda_{I/2, q}^{-1} \cong \begin{bmatrix} 1 & 0 \\
0 & \tilde{\Lambda}_{I/2, q}^{-1}
\end{bmatrix} \text{ and }\mathcal{N} \cong \begin{bmatrix} 1 & 0 \\
0 & \tilde{\mathcal{N}}
\end{bmatrix}.
\]
Furthermore we can explicitly calculate $\tilde{\Lambda}_{I/2, q}^{-1}$ and $\tilde{\mathcal{N}}$ to obtain
\[
\tilde{\Lambda}_{I/2, q}^{-1} = \frac{1}{1 - q} I \text{ and } \tilde{\mathcal{N}} = (1 - p) I + \tilde{U} \begin{bmatrix}
    p_x - p_y - p_z & 0 & 0 \\
    0 & p_y - p_x - p_z & 0 \\
    0 & 0 & p_z - p_x - p_y
\end{bmatrix} \tilde{V},
\]
where $\tilde{U}$ is a $3 \times 3$ unitary matrix with matrix elements given by $\tilde{U}_{a,a'} = \text{Tr}(A U A' U^\dagger)$ for $A, A' \in \{X, Y, Z\}$, and $\tilde{V}$ is defined similarly. Now, $\smallnorm{\mathcal{N}\Lambda^{-1}_{I/2, q}}_{2\to 2} = \text{max}(1, \smallnorm{\tilde{\Lambda}^{-1}_{I/2, q} \tilde{\mathcal{N}}})$ and
\[
\smallnorm{\tilde{\Lambda}^{-1}_{I/2, q} \tilde{\mathcal{N}}} = \frac{1}{1 - q} \smallnorm{\tilde{N}} \leq \frac{1}{1 - q}\bigg(1 - p + \text{max}(p_x - p_y - p_z, p_y - p_x - p_z, p_z - p_x - p_y)\bigg).
\]
Now, we clearly have that $\text{max}(p_x - p_y - p_z, p_y - p_x - p_z, p_z - p_x - p_y) \leq p - \text{min}(p_x, p_y, p_z)$ and thus $\smallnorm{\tilde{\Lambda}^{-1}_{I/2, q} \tilde{\mathcal{N}}} \leq (1 - \text{min}(p_x, p_y, p_z)) / (1 -q)$. Thus, choosing $q = \text{min}(p_x, p_y, p_z)$ yields that $\smallnorm{\Gamma_{I/2}^{-1/2} \mathcal{N}\Lambda_{\sigma, p}^{-1}\Gamma_{I/2}^{1/2}}_{2\to 2} = \smallnorm{\mathcal{N}\Lambda_{I/2, q}^{-1}}_{2\to 2} \leq 1$ --- thus, we can now apply lemma \ref{lemma:hirche} with $\sigma = I/2$, $q = \min(p_x, p_y, p_z)$ which yields $\alpha = 1 - \min(p_x, p_y, p_z)$. In particular, if $\rho$ is a $N-$qubit density matrix, we obtain from lemma \ref{lemma:hirche} that
\[
D_2\bigg(\mathcal{N}^{\otimes N}(\rho) \bigg |\hspace{-0.005in}\bigg|\frac{I^{\otimes N}}{2^N}\bigg)  \leq \big(1 - \min(p_x, p_y, p_z)\big)D_2\bigg(\rho \bigg |\hspace{-0.005in}\bigg|\frac{I^{\otimes N}}{2^N}\bigg)
\]
Now consider the setting where starting from a pure state $\rho_0$, a sequence of $N$-qubit unitaries $U_1, U_2 \dots U_d$ is applied interspersed with the noise channel $\mathcal{N}$ acting on each qubit. Since $D_2(U_i \rho U_i^\dagger |\hspace{-0.005in}|(I/2)^{\otimes N}) = D_2( \rho  |\hspace{-0.005in}|(I/2)^{\otimes N})$, the final state $\rho_d = \mathcal{N}^{\otimes N} \mathcal{U}_d \mathcal{N}^{\otimes N} \mathcal{U}_{d - 1} \dots \mathcal{N}^{\otimes N} \mathcal{U}_1(\rho_0) $ satisfies
\[
D_2\bigg(\rho_d \bigg |\hspace{-0.005in}\bigg|\frac{I^{\otimes N}}{2^N}\bigg)  \leq \big(1 - \min(p_x, p_y, p_z)\big)^dD_2\bigg(\rho_0 \bigg |\hspace{-0.005in}\bigg|\frac{I^{\otimes N}}{2^N}\bigg) = N(1 - \min(p_x, p_y, p_z))^d.
\]
This completes the proof. \hfill $\square$

\emph{Non-unital noise channels}. Next, we consider non-unital noise channels on $\mathbb{C}^d$ (where we are typically interested in $d = 2$) --- we will restrict ourselves to noise channels which have a full Kraus rank i.e.~the Kraus operators $K_1, K_2 \dots K_{d^2}$ span the entire space of operators on $\mathbb{C}^d$. Given a noise rate $p$, we will assume that the noise channel $\mathcal{N}$ is given by
\begin{align}\label{eq:primitive}
\mathcal{N}(X) = (1 - p) X + p \sum_{i = 1}^{d^2}K_i X K_i^\dagger,
\end{align}
where $\sum_{i = 1}^{d^2}K_i^\dagger K_i = I$. A common example of such a channel would be $\mathcal{N}(X) = (1 - p) X + p \text{Tr}(X) \tau$, where $\tau \succ 0$ i.e.~a channel that traces the qudit and replaces it with a, generally non-identity, full rank state. Physically, this would be a good model for an environment that disentangles the qubits in the quantum circuit, and brings them to a finite temperature state.

For channels of the form of Eq.~\ref{eq:primitive}, the following lemma straightforwardly follows.
\begin{lemma}\label{lemma:decomposition}
    Suppose $\mathcal{N}$ is a channel of the form given in Eq.~\ref{eq:primitive} with unique fixed point $\tau \succ 0$, $\exists$ another channel $\mathcal{N}'$ with fixed point $\tau$ such that $\mathcal{N}(X) = \Lambda_{\tau, q}  \mathcal{N}'(X) $, where $\Lambda_{\tau, q}(X) = (1 - q) X + q\textnormal{Tr}(X) \tau$, for some $q \in (0, 1)$. Furthermore, the largest such $q$ can be computed by solving the semi-definite program
    \begin{equation}\label{eq:}
   \begin{aligned}
\max_{q} \quad & q\\
\textnormal{s.t.} \quad & \Phi_\mathcal{N} - q I\otimes \tau \succeq 0,\\
  &q > 0, q < 1.    \\
\end{aligned},
\end{equation}
where $\Phi_\mathcal{N}$ is the Choi state of $\mathcal{N}$.
\end{lemma}
\noindent\emph{Proof}: Since $\mathcal{N}$ has a full Kraus rank, we can find linearly independent operators $L_1, L_2 \dots L_{d^2}$ such that $\sum_{i = 1}^{d^2} L_i^\dagger L_i = I$ and $\mathcal{N}(X) = \sum_{i = 1}^{d^2} L_i X L_i^\dagger$. Furthermore, since $L_1, L_2 \dots L_{d^2}$ are linearly independent, it follows that for any $M \in \mathbb{C}^{d\times d}$, $\text{Tr}(L_i^\dagger M) = 0 \ \forall \ i \in \{1, 2 \dots d^2\}$ implies that $M = 0$. Consequently, $\exists \lambda_0 > 0$ such that
\begin{align}\label{eq:min_singular_val}
\forall M \in \mathbb{C}^{d\times d}: \sum_{i = 1}^{d^2} \abs{\text{Tr}(L_i^\dagger M)}^2 \geq \lambda_0 \text{Tr}(M^\dagger M).
\end{align}
Now, suppose $q \in (0, 1)$, and $\mathcal{N}'= (\mathcal{N} - q\textnormal{Tr}(\cdot) \tau)/ (1 - q)$. It is clear that $\mathcal{N}'$ is trace preserving. We need to establish that $\mathcal{N}'$ is also completely positive. For that, consider the Choi State corresponding to $\mathcal{N}'$, $\Phi_{\mathcal{N}'}$:
\begin{align*}
\Phi_{\mathcal{N}'} &= (\mathcal{N}'\otimes \textnormal{id})(\ket{\Phi}\!\bra{\Phi}), \nonumber \\
&=\frac{1}{1 - q}\bigg( \sum_{i = 1}^{d^2} \sum_{j, k = 1}^d L_i\ket{j}\!\bra{k}L_i^\dagger \otimes \ket{j}\!\bra{k} - q I \otimes \tau \bigg), \nonumber \\
&=\frac{1}{1 - q}\bigg( \sum_{i = 1}^{d^2} \sum_{j, k, j', k' = 1}^d (L_i)_{j', j} (L_i)_{k', k}^*\ket{j'}\!\bra{k'} \otimes \ket{j}\!\bra{k} - q I \otimes \tau \bigg).
\end{align*}
For $\mathcal{N}'$ to be completely positive, it is necessary and sufficient for $\Phi_{\mathcal{N}'} \succeq 0$. To impose this condition, consider a state $\ket{\psi} \in \mathbb{C}^{d}\otimes \mathbb{C}^d$, then
\begin{align*}
\bra{\psi}\Phi_{\mathcal{N}'}\ket{\psi} &= \frac{1}{1 - q}\bigg( \sum_{i = 1}^{d^2} \sum_{j, k, j', k' = 1}^d (L_i)_{j', j} (L_i)_{k', k}^*\psi^*_{j', j}\psi_{k', k}- q \bra{\psi} I \otimes \tau \ket{\psi}\bigg), \nonumber\\
&= \frac{1}{1 - q}\bigg( \sum_{i = 1}^{d^2}  \abs{\text{Tr}(L_i^\dagger \Psi)}^2 - q \bra{\psi} I \otimes \tau \ket{\psi}\bigg),
\end{align*}
where $\Psi \in \mathbb{C}^{d\times d}$ is the state $\ket{\psi}$ reshaped as a matrix. Using Eq.~\ref{eq:min_singular_val}, we obtain that
\[
\bra{\psi}\Phi_{\mathcal{N'}}\ket{\psi} \geq \frac{1}{1 - q}\bigg(\text{Tr}(\Psi^\dagger \Psi) \lambda_0 - q\norm{\tau} \norm{\ket{\psi}}^2\bigg) = \frac{\lambda_0 - q\norm{\tau}}{1 - q} \norm{\ket{\psi}}^2.
\]
Thus, for $q = \lambda_0 / \norm{\tau}$, we obtain that $\Phi_{\mathcal{N}'} \succeq 0$ and hence $\mathcal{N}'$ is completely positive. This establishes that there exists $q > 0$ such that $\mathcal{N} = (1 - q) \mathcal{N}' + q \textnormal{Tr}(\cdot) \tau$. Furthermore, it trivially follows that $\mathcal{N}'(\tau) = \tau$ from this definition of $\mathcal{N}'$. It therefore also follows that $\mathcal{N} = \Lambda_{\tau, q} \mathcal{N}'$. The optimization problem written for the calculation of $q$ in Eq.~\ref{eq:opt_problem} is simply a reformulation of the condition that the Choi state of $\mathcal{N}'$ needs to be positive semi-definite. \hfill $\square$

Finally, we now consider the setting of $N-$qubits which have unitaries $U_1, U_2 \dots U_d$ applied to them, along with noise $\mathcal{N}$ acting on each qubit at every time-step. As described in the main text, we will assume that each unitary $U_t$ is composed of two-qubit and single-qubit gates, with the gate set being chosen such that the two-qubit gates leave $\sigma^{\otimes 2}$ invariant (i.e.~the two-qubit gate is diagonal on the basis of eigenvectors of $\sigma$). \\

\noindent\emph{Proof of lemma 3}: Suppose $\rho$ is a $N-$qubit density matrix, and consider the state $\mathcal{N}^{\otimes N}(\rho)$. We will first establish that 
\begin{align}\label{eq:impact_non_unital_noise_cross_entropy}
D(\mathcal{N}^{\otimes N}(\rho) |\hspace{-0.005in}| \tau^{\otimes N}) \leq (1 - q) D(\rho |\hspace{-0.005in}| \tau^{\otimes N}) 
\end{align}
Expressing $\mathcal{N} = \Lambda_{\tau, q}\mathcal{N}'$, $\mathcal{N}^{\otimes N} =\Lambda_{\tau, q}^{\otimes N} {\mathcal{N}'}^{\otimes N} $. Suppose $\rho' = {\mathcal{N}'}^{\otimes N}(\rho)$ and then $D(\mathcal{N}^{\otimes N}(\rho) |\hspace{-0.005in}|\tau^{\otimes N}) = D(\Lambda_{\tau, q}^{\otimes N}(\rho') |\hspace{-0.005in}|\tau^{\otimes N}) $. We point out that $\Lambda_{\tau, q}$ can be viewed as generated from the Lindbladian $\mathcal{L}_{\tau} = \text{id} - \textnormal{Tr}(\cdot) \tau $ with evolution time $t = -\log(1 - q)$. It now follows from theorem 19 of Ref.~\cite{beigi2020quantum} that if $\tilde{\mathcal{L}}_{\tau} := \sum_{i = 1}^N \mathcal{L}^{(i)}_{\tau}$, where $\mathcal{L}_{\tau}^{(i)}$ is the Lindbladian $\mathcal{L}_{\tau}$ acting on the $i^\text{th}$ qubit, satisfies $D(e^{\tilde{\mathcal{L}}_{\tau}t}(\rho')|\hspace{-0.005in} | \tau^{\otimes N}) \leq e^{-t} D(\rho' |\hspace{-0.005in}|\tau^{\otimes N})$. Setting $t = -\log(1 -q)$ in this inequality, we obtain that 
$D(\Lambda_{\tau, q}^{\otimes N}(\rho')|\hspace{-0.005in}|\tau^{\otimes N}) \leq (1 - q)D(\rho'|\hspace{-0.005in}|\tau^{\otimes N})$. Now, this yields that 
\[
D(\mathcal{N}^{\otimes N}(\rho) |\hspace{-0.005in}| \tau^{\otimes N}) \leq (1 - q) D({\mathcal{N}'}^{\otimes N}(\rho) |\hspace{-0.005in}|\tau^{\otimes N}) = (1 - q)D({\mathcal{N}'}^{\otimes N}(\rho) |\hspace{-0.005in}|\mathcal{N}'(\tau^{\otimes N})) \leq (1 - q)D(\rho |\hspace{-0.005in}|\tau^{\otimes N}).
\]
Next, we consider one of the unitaries applied in the circuit, $U_t$. The two-qubit gates in this unitary, by assumption, leave $\tau^{\otimes 2}$ unchanged. Thus, if $V^{(2)}$ is a two-qubit unitary that satisfies this condition, we have that
\[
D(V^{(2)}\rho {V^{(2)}}^\dagger |\hspace{-0.005in}| \tau^{\otimes N}) = D(\rho |\hspace{-0.005in}|{V^{(2)}}^\dagger \tau^{\otimes N}V^{(2)}) = D(\rho |\hspace{-0.005in}| \tau^{\otimes N}).
\]
Furthermore, if $V^{(1)}$ is a single qubit gate, we obtain from the data processed triangle inequality that
\[
D(V^{(1)}\rho {V^{(1)}}^{\dagger} |\hspace{-0.005in}| \tau^{\otimes N}) \leq D(\rho |\hspace{-0.005in}| \tau^{\otimes N}) + D_\infty({V^{(1)}}\tau {V^{(1)}}^{\dagger} |\hspace{-0.005in}|\tau).
\]
Suppose that the single qubits applied in $U_t$ are $V_{t, \alpha}^{(1)}$ for $\alpha \in \{1, 2, 3 \dots\}$, then from the above two inequalities we obtain that
\begin{align}\label{eq:unitary_impact}
D(U_t \rho U_t^\dagger |\hspace{-0.005in}|\tau^{\otimes N}) \leq D(\rho |\hspace{-0.005in}|\tau^{\otimes N}) + \sum_{\alpha} D_\infty({V_{t, \alpha}^{(1)}}\tau {V_{t, \alpha}^{(1)}}^{\dagger}|\hspace{-0.005in}|\tau).
\end{align}
Recursing Eqs.~\ref{eq:impact_non_unital_noise_cross_entropy} and \ref{eq:unital_primitive}, we obtain that $\rho_t = U_t U_{t - 1} \dots U_1 \rho_0 U_1^\dagger \dots U_{t - 1}^\dagger U_t^\dagger$ satisfies
\[
D(\rho_t |\hspace{-0.005in}|\tau^{\otimes N}) \leq (1 - q)^t D(\rho_0 |\hspace{-0.005in}|\tau^{\otimes N}) + \sum_{s = 0}^{t - 1} \sum_{\alpha}(1 - q)^{t - s} D_\infty({V_{s + 1, \alpha}^{(1)}}\tau {V_{s + 1, \alpha}^{(1)\dagger}}|\hspace{-0.005in}|\tau)
\]
Finally, this bound can be translated to a bound on $\norm{\rho_t - \sigma^{\otimes N}}_F$ by a standard application of Pinsker's inequality as shown
\[
\norm{\rho_t - \tau^{\otimes N}}_F^2 \leq 2\norm{\rho_t - \tau^{\otimes N}}_1 \leq \sqrt{2 D(\rho_t |\hspace{-0.005in}|\tau^{\otimes N})},
\]
which completes the proof of the lemma. \hfill$\square$

\section{Free fermion formulation}

In this section, we describe our approach to numerical studies of Gaussian fermionic systems. We describe the fermionic system model and how the energy of the state at the output of the noisy circuit and the information content-based dual function \eqref{eq:ic_dual_function} can be efficiently calculated by utilizing covariance matrices and quadratic Hamiltonian transformations. 

\subsection{Model}
We consider $N$ fermions arranged in a lattice and define for each lattice site $x$ the Majorana operators $c_x^1, c_x^2$ as satisfying the following anti-commutation relations, 
\begin{align*} \label{eq:majoranaanticomm}
    &\{ c^1_x, c^1_{x'}\} = \{ c^2_x, c^2_{x'}\} = \delta_{xx'}\\
    & \{ c^1_x, c^2_{x'}\} =  0, \ \forall x, x' \in \text{lattice}. 
    \numberthis
\end{align*} 
We choose the target Hamiltonian $H$ to be quadratic, 
\[
H = i\sum_{\substack{\alpha, \alpha', x, x'}} h_{x, x'}^{\alpha, \alpha'} c_x^\alpha c_{x'}^{\alpha'}. 
\] Defining the operator valued vector, 
\begin{align*}
    \vec{c} = \begin{pmatrix}
    c^1_1, & c^1_2, & \dots, & 
    c^1_N, &
    c^2_1, &
    c^2_2, &
    \dots, & 
    c^2_N
    \end{pmatrix}^T,
    \numberthis
\end{align*} 
the target Hamiltonian can be written as $H = i \vec{c}^{\ T} h \vec{c}$ where the matrix $h$ specifying the Hamiltonian $H$ is real and anti-symmetric. The ground state energy of the target Hamiltonian can be computed by diagonalization of its compact representation $h$ \cite{suraceFermionicGaussianStates2022a}. 

\subsection{Calculation of the noisy circuit output}

The output energy, with respect to the quadratic target Hamiltonian $H$, can be computed from the covariance matrix of the state at the end of the circuit. We define the covariance matrix of a state $\rho$ as,
\begin{align*}
    \gamma^{\alpha, \alpha'}_{x, x'} = i \text{Tr}(\rho [c^{\alpha}_x, c^{\alpha'}_{x'}]),
    \numberthis
\end{align*} 
where $[.,.]$ denotes the commutator. The energy of a state $\rho$ with respect to the target Hamiltonian $H$ is then \cite{suraceFermionicGaussianStates2022a}, 
\begin{align*}
    \text{Tr}(\rho H) = - \frac{\text{Tr}(h \gamma)}{2}. 
    \numberthis
\end{align*} 
The covariance matrix of the state after each instance of Gaussian unitary or noise in the circuit can be calculated by evolving the covariance matrix $\gamma_0$ of the vacuum ($\gamma_0$ can be analytically calculated \cite{suraceFermionicGaussianStates2022a}) using the following transformations, 

\begin{enumerate}
    \item \emph{Unitary:} Any Gaussian unitary can be expressed as $U = \exp(-i H_U)$, where $H_U = i \vec{c}^{\ T} h_U \vec{c}$ is the generating quadratic Hamiltonian, and it transforms the covariance matrix of the state as \cite{suraceFermionicGaussianStates2022a},
    \begin{align*}
        \gamma \rightarrow e^{2 h_U} \gamma e^{-2 h_U}. 
        \numberthis
    \end{align*}
    
    \item \emph{Noise:} Depolarizing noise with probability $p$ acting on the fermion at lattice site $x$ transforms the covariance matrix in the following manner \cite{suraceFermionicGaussianStates2022a, schuchMatrixProductState2019},
\begin{align*}
    \gamma \rightarrow p \tilde{\gamma} + (1-p) \gamma
    \numberthis,
\end{align*} where the matrix $\tilde{\gamma}$ is obtained from the input covariance matrix $\gamma$ by zeroing out the rows and columns corresponding to the fermion at site $x$, i.e., 
\begin{align*}
    \tilde{\gamma}^{\alpha, \alpha'}_{x', x''} = \begin{cases}
    0, & \text{if } x' = x \ \text{or} \ x'' = x \\
    \gamma^{\alpha, \alpha'}_{x', x''}, & \text{else}. 
    \end{cases}
    \numberthis
\end{align*}
\end{enumerate} 

\subsection{Calculation of the dual function}

We choose the dual variables ${\sigma_1, \sigma_2, \ldots, \sigma_d}$ to be quadratic Hamiltonians, i.e. we consider the following ansatz, 
\[
\sigma_t = i\sum_{\substack{\alpha, \alpha', x, x' \\ d(x, x') \leq r}} s_{x, x'; t}^{\alpha, \alpha'} c_x^\alpha c_{x'}^{\alpha'} = i \vec{c}^{\ T} s_t \vec{c} , \ t \in \{1, \ldots, d\},
\] where $d(x, x')$ is a distance measure between two lattice sites and the coefficients $s_{x, x'; t}^{\alpha, \alpha'}$ are real and can be arranged into an anti-symmetric matrix $s_t$. $\sigma_t$ are thus local Hamiltonians with interaction range $r$. 

In order to evaluate the circuit dual function \eqref{eq:ic_dual_function}, we calculate the effect of the circuit channels on the dual variables first, i.e. $\mathcal{E}_t^{\dagger}(\sigma_t)$. Each such channel is the composition of a Gaussian unitary channel and the depolarizing channel and both the unitary and noise channels map quadratic operators to quadratic operators. Hence, the channel's action can be expressed as a transformation of the compact Majorana representation $s_t$ of the dual variable $\sigma_t$ using the following transformation rules,  
\begin{enumerate}
    \item \emph{Unitary:} Any Gaussian unitary can be expressed as $U = \exp(-i H_U)$, where $H_U = i \vec{c}^{\ T} h_U \vec{c}$ is the generating quadratic Hamiltonian. Conjugation of the dual variable $\sigma_t = i \vec{c}^{\ T} s_t \vec{c}$ by $U$ transforms it into another quadratic operator $\tilde{\sigma}_t = U \sigma_t U^{\dagger} = i \vec{c}^{\ T} \tilde{s}_t \vec{c} $, with the transformed Majorana representation $\tilde{s}_t$, 
    \begin{align*}
        \tilde{s}_t = e^{2 h_U} s_t e^{-2 h_U}. 
        \numberthis
    \end{align*} This transformation can be derived by expanding the conjugation using the Baker-Campbell-Hausdorff formula $e^B A e^{-B} = A + [B, A] + \frac{1}{2!} [B, [B, A]] + \frac{1}{3!} [B, [B, [B, A]]] + \ldots $, and using the commutation relation,
    \begin{align*}
        [c^{\alpha_1}_{x_1} c^{\alpha_2}_{x_2}, c^{\alpha_3}_{x_3} c^{\alpha_4}_{x_4}] =& - c^{\alpha_1}_{x_1} c^{\alpha_3}_{x_3} \delta_{\alpha_2, \alpha_4} \delta_{x_2, x_4} + c^{\alpha_1}_{x_1} c^{\alpha_4}_{x_4} \delta_{\alpha_2, \alpha_3} \delta_{x_2, x_3} \\ 
        &- c^{\alpha_3}_{x_3} c^{\alpha_2}_{x_2} \delta_{\alpha_1, \alpha_4} \delta_{x_1, x_4} + c^{\alpha_4}_{x_4} c^{\alpha_2}_{x_2} \delta_{\alpha_1, \alpha_3} \delta_{x_1, x_3}. 
        \numberthis
    \end{align*}
    
    \item \emph{Noise:} The noise channel $\mathcal{N}_x$ corresponding to depolarizing noise on the fermion at site $x$ with probability $p$ transforms the dual variable $\sigma_t$ as, 
    \begin{align*}
        \mathcal{N}_x (\sigma_t) = (1 - p) \sigma_t + p \text{Tr}_x (\sigma_t) \otimes \frac{\mathds{1}}{2},
        \numberthis
    \end{align*} where $\text{Tr}_x(.)$ denotes partial trace over fermion at site $x$. Expanding out the partial trace term using linearity, 
    \begin{align*} \label{eq:zt}
        \text{Tr}_x (\sigma_t) \otimes \frac{\mathds{1}}{2} &= \sum_{\alpha', \alpha'', x', x''} i s_{x', x''; t}^{\alpha', \alpha''} \text{Tr}_x (c_{x'}^{\alpha'} c_{x''}^{\alpha''}) \otimes \frac{\mathds{1}}{2}, \\
        &= \sum_{\alpha', \alpha'', x', x''} i s_{x', x''; t}^{\alpha', \alpha''} (1 - \delta_{x, x'}) (1 - \delta_{x, x''}) c_{x'}^{\alpha'} c_{x''}^{\alpha''},
        \numberthis
    \end{align*} where we obtain the second line by using $\text{Tr}(c_{x'}^{\alpha'}) = 0 $ and $s_{x', x'; t}^{\alpha', \alpha'} = 0$ $\forall (x', \alpha')$. Thus, the noise channel transforms the dual variable into another quadratic operator $\tilde{\sigma}_t = i \vec{c}^{\ T} \tilde{s}_t \vec{c} $ with the transformed Majorana representation, 
    \begin{align*}
        \tilde{s}_t = p s_t + (1-p) z_t
        \numberthis, 
    \end{align*} 
    where, using \eqref{eq:zt}, the matrix $z_t$ is obtained from $s_t$ by zeroing out the rows and columns corresponding to the fermion at site $x$, i.e., 
    \begin{align*}
        z^{\alpha', \alpha''}_{x', x''; t} = \begin{cases}
        0, & \text{if } x' = x \ \text{or} \ x'' = x \\
        s^{\alpha', \alpha''}_{x', x''; t}, & \text{else}. 
        \end{cases}
        \numberthis
    \end{align*}
\end{enumerate} 
To calculate the dual function \eqref{eq:ic_dual_function}, we need to calculate the Gibbs free energy of $H_t = \sigma_t - \mathcal{E}_{t+1}^{\dagger}(\sigma_{t + 1})$ where $H_t$ is a quadratic operator itself because, as mentioned earlier, the channel $\mathcal{E}^{\dagger}_{t + 1}$ maps quadratic operators to quadratic operators. Therefore, $H_t = i \vec{c}^{\ T} h_t \vec{c} $ where the Majorana representation $h_t$ can be obtained using the transformations described above. We calculate the free energy by diagonalizing $H_t$. This can be achieved by diagonalizing $h_t$ --- as $h_t$ is anti-symmetric, it is possible \cite{suraceFermionicGaussianStates2022a} to find an orthogonal transformation $O$ such that,
\begin{align*}
    O^T h_t O = \bigoplus_{\alpha, x} \begin{pmatrix} 
    0 & \epsilon_{x; t} \\
    -\epsilon_{x; t} & 0. 
    \end{pmatrix}
\end{align*} We define the operator valued vector $\vec{\tilde{c}} = O^T \vec{c} = (\tilde{c}^1_1, \tilde{c}^1_2, \ldots, \tilde{c}^1_N, \tilde{c}^2_1, \tilde{c}^2_2, \ldots, \tilde{c}^2_N)^T$ and, due to the orthogonality of $O$, the operators $\tilde{c}^{\alpha}_x$ satisfy the Majorana anti-commutation relations \eqref{eq:majoranaanticomm}. Using the newly defined Majorana operators, $H_t$ can be written in a diagonal form as, 
\begin{align*}
    H_t = i \sum_{x} \epsilon_{x; t} \left( \tilde{c}^1_x \tilde{c}^2_x - \tilde{c}^2_x \tilde{c}^1_x \right). 
\end{align*} Using the diagonal form, the Gibbs free energy for $H_t$ is,
\begin{align*}
    \mathcal{F}(H_t, \lambda_t) &= \log\text{Tr}\exp\left(-\frac{i}{\lambda_t} \sum_{x} \epsilon_{x; t} \left( \tilde{c}^1_x \tilde{c}^2_x - \tilde{c}^2_x \tilde{c}^1_x \right) \right) \\
    &= \log \prod_{x} \text{Tr}\exp\left(-\frac{i \epsilon_{x; t}}{\lambda_t}  \left( \tilde{c}^1_x \tilde{c}^2_x - \tilde{c}^2_x \tilde{c}^1_x \right) \right).
\end{align*}
The operator exponentials above can be represented in the Fock space basis of the fermion at site $x$,
\begin{align*}
   \exp\left(-\frac{i \epsilon_{x; t}}{\lambda_t}  \left( \tilde{c}^1_x \tilde{c}^2_x - \tilde{c}^2_x \tilde{c}^1_x \right) \right) = \begin{pmatrix} 
   e^{\epsilon_{x;t}/\lambda_t} & 0 \\
   0 & e^{-\epsilon_{x;t}/\lambda_t}.
   \end{pmatrix}
\end{align*} The free energy then simplifies to,
\begin{align*}
    \mathcal{F}(H_t, \lambda_t) = \sum_x \log(e^{\epsilon_{x;t}/\lambda_t} + e^{-\epsilon_{x;t}/\lambda_t}),
\end{align*} and therefore it can be calculated just from the diagonalization of the $2N \times 2N$ dimensional matrix $h_t$.

\bibliography{vqa_paper}{}

\begin{thebibliography}{57}
\bibitem [{not()}]{notation}%
  \BibitemOpen
  \href@noop {} {}\bibinfo {note} {Given a finite-dimensional Hilbert space $\mathcal{H}$, we use $\mathcal{D}_1(\mathcal{H})$ to denote the set of all density matrices on $\mathcal{H}$, and $\mathcal{M}(\mathcal{H})$ to denote the set of Hermitian linear operators on $\mathcal{H}$. We use the computer-science big-O notation for function asymptotics. In particular, given two functions $f, g:[0, \infty) \to [0, \infty)$, $f = O(g)$ if, for some $c > 0$, $f(x) \leq c g(x)$ as $x \to \infty$ and $f = \Omega(g)$ if, for some $c > 0$, $f(x) \geq c g(x)$ as $x \to \infty$.}\BibitemShut {Stop}%
\end{thebibliography}

\end{document}